\documentclass[twocolumn,showpacs,preprintnumbers,amsmath,amssymb]{revtex4}
\usepackage{epsf}

\def\e{\mathrm{e}}
\def\cc{cosmological constant}

\def\be{\begin{equation}}
\def\ee{\end{equation}}
\def\bea{\begin{eqnarray}}
\def\eea{\end{eqnarray}}
\def\d{\mbox{d}}
\def\e{\mathop{\rm e}}

\def\min{\mathop{\rm min}}
\def\max{\mathop{\rm max}}
\def\p{\partial}
\def\oder#1#2{\frac{\d #1}{\d #2}}
\def\pder#1#2{\frac{\partial #1}{\partial #2}}
\def\ra{\rightarrow}
\let\phi=\varphi

\def\Sds{Schwarzschild--de~Sitter }
\def\RNds{Reissner--Nordstr\"{o}m--de Sitter }
\def\RNads{Reissner--Nordstr\"{o}m--(anti-)de~Sitter }
\def\Kads{Kerr--(anti-)de Sitter }
\def\Kds{Kerr--de~Sitter }

\begin{document}


\title{Equatorial circular orbits in the Kerr--de~Sitter spacetimes}
\author{Zden\v{e}k Stuchl\'{\i}k}
  \email{Zdenek.Stuchlik@fpf.slu.cz}
\author{Petr Slan\'{y}}
  \email{Petr.Slany@fpf.slu.cz}
\affiliation{Department of Physics, Faculty of Philosophy and Science,
             Silesian University at Opava\\ 
             Bezru\v{c}ovo n\'{a}m. 13, 746 01 Opava, Czech Republic}

\date{received 18 June 2003; published 1 March 2004}

\begin{abstract}
Equatorial motion of test particles in Kerr--de~Sitter spacetimes is
considered. Circular orbits are determined, their properties are
discussed for both black-hole and naked-singularity spacetimes, and their
relevance for thin accretion disks is established. The circular orbits 
constitute two families that coalesce at the so-called static radius. The orientation
of the motion along the circular orbits is, in accordance with case of  
asymptotically flat Kerr spacetimes, defined by relating the motion to the locally 
nonrotating frames. The minus-family orbits are all counterrotating, while 
the plus-family orbits are usually corotating relative to these frames. 
However, the plus-family orbits become counterrotating in the vicinity of the 
static radius in all Kerr--de~Sitter spacetimes, and they become 
counterrotating in the vicinity of the ring singularity in Kerr--de~Sitter 
naked-singularity spacetimes with a low enough rotational parameter. 
In such spacetimes, the efficiency of the conversion of the rest energy into heat 
energy in the geometrically thin plus-family accretion disks can reach extremely high 
values exceeding the efficiency of the annihilation process. The transformation of
a Kerr--de~Sitter naked singularity into an extreme black hole due to accretion
in the thin disks is briefly discussed for both the plus-family and
minus-family disks. It is shown that such a conversion leads to 
an abrupt instability of the innermost parts of the plus-family accretion
disks that can have strong observational consequences.
\end{abstract}


\maketitle

\section{Introduction}

It is commonly accepted that the energy sources of quasars and active galactic nuclei
are accretion disks around central massive black holes \cite{Abr-Per:1997:CLAQG:,%
                                                             Bla:1990:AGN:}.
The basic properties of geometrically thin accretion disks (with negligible pressure) are
determined by the circular geodesic motion in the black-hole backgrounds 
\cite{Nov-Tho:1973:BlaHol:}. The basic properties of geometrically thick disks are determined
by the equilibrium configurations of a perfect fluid orbiting in the black-hole background;
however, the geodesic structure of the background is relevant also for the properties
of the thick disks \cite{Jar-Abr-Pac:1980:ACTAS:}.

According to the cosmic censorship hypothesis \cite{Pen:1969:NUOC2:} and the
uniqueness theorems for black holes 
\cite{Car:1973:BlaHol:}, the result of the gravitational collapse of a sufficiently massive 
rotating object is a rotating Kerr black hole, rather than a Kerr naked singularity; further, 
the laws of
black-hole thermodynamics forbid conversion of black holes into a naked singularity. However, 
although the cosmic censorship is a very plausible hypothesis, there is some evidence against 
it. Naked singularities arise in various models of spherically symmetric collapse (e.g. 
\cite{Lak-Zan:1990:PHYSR4:}). In modeling the collapse of rotating stars, it
was pointed out that,
although mass shedding and gravitational radiation will reduce the angular momentum of the 
star during collapse, it will not in general be reduced to the value
that corresponds to a Kerr black hole \cite{Mil-deFel:1985:ASTRJ2:}.The 2D
numerical models \cite{Nak-Ooh-Koj:1987:PROTP3:} imply that a rotating,
collapsing supermassive object will not always dissipate enough
angular momentum to form a Kerr black hole, but a Kerr-like naked
singularity has to be expected to develop from objects rotating
rapidly enough. Candidates for the formation of naked Kerr geometry
with a ring singularity from the collapse of rotating stars were found in the
scenario of \cite{Char-Cla:1990:CLAQG:}. The numerical models of the
collapse of collisionless gas spheroids also results in strong
candidates for the formation of naked singularities
\cite{Sha-Teu:1992:PHYSR4:}. 

Because Penrose's cosmic censorship hypothesis \cite{Pen:1969:NUOC2:} is far from being 
proved, naked singularity spacetimes related to black-hole spacetimes with a nonzero
charge and/or rotational parameter could still be considered conceivable models of 
quasars and active galactic nuclei and deserve some attention. Of particular interest
are those effects that could distinguish a naked singularity from black holes.

Test particle motion and test fields were extensively studied for
Kerr black-hole spacetimes \cite{Car:1968:PHYSREV:,%
                                 Bar-Pre-Teu:1972:ASTRJ2:,%
                                 Bar:1973:BlaHol:,%
                                 Ste-Wal:1973:SpringerTracts:,%
                                 Bic-Stu:1976:BULAI:,%
                                 Sha:1979:GENRG1:,%
                                 Con:1984:GENRG1:,%
                                 Dym:1986:SOVPU2:,%
                                 Bic-Stu-Bal:1989:BULAI:,%
                                 Bal-Bic-Stu:1989:BULAI:}. 
Gravitational radiation of
particles moving in the field of a Kerr black hole were discussed in
\cite{Sai-Shi-Mae:1997:PHYSR4:}; the motion of spinning test particles was discussed in
\cite{Sem:1999:MONNR:}. For a detailed review see the books of
Chandrasekhar \cite{Chan:1983:BlackHoles:} and Frolov and Novikov  
\cite{Fro-Nov:1998:BlackHolePhys:}. However, Kerr naked singularities were also
studied widely. Their repulsive effects and causality-violating
regions were investigated by de Felice and co-workers
\cite{deFel:1975:ASTRA:,%
      deFel-Cal:1979:GENRG1:,%
      deFel-Bra:1988:CLAQG:},
the equatorial circular geodesics and motion of spherical shell of
incoherent dust were investigated in \cite{Stu:1980:BULAI:}, the
collimation effect of the region nearby the ring singularity was treated in
\cite{Bic-Sem-Had:1993:MONNR:}, and the motion of spinning test particles was discussed in
\cite{Sem:1999:MONNR:}. Chandrasekhar \cite{Chan:1983:BlackHoles:}
also devotes some attention to the effects of Kerr naked singularities,
saying ``considerable interest attaches to knowing the sort of things
space-times with naked singularities are and whether there are any
essential differences in the manifestations of space-times with
singularities concealed behind event horizons.'' We follow Chandrasekhar's approach. 
  
All recently available data from a wide variety of cosmological tests indicate 
convincingly that in the framework of the inflationary cosmology a nonzero, 
although very small, repulsive \cc\ $\Lambda > 0$ has to be invoked in 
order to explain the dynamics of the recent Universe 
\cite{Bah-etal:1999:SCIEN:,%
      Kol-Tur:1990:EarUni:}. 
The presence of a repulsive \cc\ changes
substantially the asymptotic structure of the black-hole (or naked-singularity) 
backgrounds, as they become asymptotically de~Sitter spacetimes, not flat spacetimes.
In such spacetimes, an event horizon always exists behind which the geometry is
dynamic; we call it a cosmological horizon. Therefore, it is relevant to clarify the
influence of the repulsive cosmological constant on the astrophysically interesting 
properties of the black-hole or naked-singularity background. For these purposes, analysis
of the geodesic motion of test particles and photons is among the most important techniques.
(It could be noted that the optical reference geometry introduced by Abramowicz and  
co-workers \cite{Abr-Car-Las:1988:GENRG1:}
reflects in an illustrative and intuitive way some hidden properties of the
geodesic motion \cite{Abr-Pra:1990:MONNR:,%
                        Hle:2002:JB60:,%
                        Stu-Hle:2000:CLAQG:}.) Of particular
interest are circular geodesics being relevant for the accretion disks.

Properties of the geodesic motion in the Schwarzschild--(anti-)de~Sitter
and Reissner--Nordstr\"{o}m-- (anti-)de~Sitter spacetimes were discussed in
\cite{Stu-Hle:1999:PHYSR4:,%
      Stu-Hle:2002:ACTPS2:}.
Properties of the circular orbits of test particles show that due to the presence
of a repulsive \cc\, the thin disks have not only an inner edge determined
(approximately) by the radius of the innermost stable circular orbit, but
also an outer edge given by the radius of the outermost stable circular orbit, 
located nearby the so-called static radius, where the gravitational attraction of 
a black hole (naked singularity) is just compensated for by the cosmological repulsion.

A similar analysis of the equilibrium configurations of a perfect fluid orbiting
the Schwarzschild--de~Sitter black-hole backgrounds allowing the existence of stable circular
orbits, which is a necessary condition for the existence of accretion disks, shows 
that also thick accretion disks have both the inner and outer
edges located nearby the inner (outer) marginally bound circular geodesic. The accretion 
through the inner cusp and the outflow of matter through the outer cusp of the equilibrium 
configurations are driven by the Paczy\'{n}ski mechanism. It is a mechanical nonequilibrium
process when the matter of the disk slightly overfills the critical equipotential surface
with two cusps and thus violates the hydrostatic equilibrium
\cite{Stu-Sla-Hle:2000:ASTRA:}.

In the case of \RNads\ backgrounds \cite{Stu-Hle:2002:ACTPS2:}, the discussion has been enriched
for the case of the naked-singularity spacetimes---it was shown that even two separated regions of 
stable circular orbits are allowed for the naked-singularity spacetimes with spacetime 
parameters appropriately chosen.

However, it is very important to understand the role of a nonzero \cc\ in the
astrophysically most relevant, rotating, Kerr backgrounds. Equatorial
motion of photons has been studied extensively for Kerr--Newman--(anti-)de 
Sitter spacetimes describing both black holes and naked singularities and
some unusual effects have been found (for details, see \cite{Stu-etal:1998:PHYSR4:,%
      Stu-Hle:2000:CLAQG:}).
Here, attention will be focused on the circular equatorial motion of test particles 
in the Kerr--de~Sitter backgrounds.

In Sec. \ref{KdS spacetimes}, the Kerr--de~Sitter backgrounds are
separated in the parameter space into black-hole and naked-singularity
spacetimes. In Sec. \ref{Eqmo}, the equations of the equatorial motion of 
test particles are presented. In Sec. \ref{Eco}, the constants of motion 
of the circular orbits are determined, and their properties are discussed. 
As in Kerr spacetimes, there exist two different sequences of the 
equatorial circular geodesics. We call them plus (minus-) family orbits.
All minus-family orbits are counterrotating relative to the locally 
nonrotating frames (LNRF; for a definition of these frames see \cite{Bar:1973:BlaHol:}),
while the plus-family orbits are mostly corotating, but in some regions are
counterrotating relative to the LNRF. Only outside the outer horizon of the Kerr
black holes are all the plus-family orbits corotating relative to the LNRF. On 
the other hand, in vicinity of the ring singularity of the Kerr naked singularities
with rotational parameter low enough, the plus-family orbits become counterrotating
relative to the LNRF \cite{Stu:1980:BULAI:}. We shall see that in all Kerr--de~Sitter 
spacetimes this happens nearby the so-called static radius, where the sequences of 
plus-family and minus-family orbits coalesce. [Note that in asymptotically flat
Kerr spacetimes, the orbits corotating (counterrotating) relative to the LNRF also
corotate (counterrotate) from the point of view of stationary observers at infinity.
However, the last criterion cannot be used in the asymptotically de~Sitter spacetimes
under consideration.] In Sec. \ref{Dis}, the properties of the circular orbits are 
discussed with attention focused on their relevance for thin accretion disks. 
Regions where the orbits of the plus-family are counterrotating relative to the LNRF
are determined; further, it is established where these orbits could have a negative 
energy parameter. The efficiency of the accretion process in geometrically thin disks 
is determined. In Sec. \ref{Conclusion}, concluding remarks are presented.

\section{Kerr--de~Sitter black-hole and~naked-singularity spacetimes}
\label{KdS spacetimes}

In the standard Boyer-Lindquist coordinates ($t,r,\theta, \phi$) and
geometric units ($c=G=1$), the \Kads geometry is given by the line element
\begin{eqnarray}
               \d s^2 & = & -\frac{\Delta_r}{I^2 \rho^2}(\d t-a\sin^2 \theta
                \d\phi)^2
                                                        \nonumber \\
                & & + \frac{\Delta_{\theta}\sin^2 \theta}{I^2 \rho^2}
                \left[a\d t- \left(r^2+a^2 \right) \d\phi \right]^2
                                                        \nonumber \\
                & & + \frac{\rho^2}{\Delta_r} \d r^2 +
                \frac{\rho^2}{\Delta_{\theta}} \d\theta^2,      \label{e1}
\end{eqnarray}
where
\bea
        \Delta_r & = & -\frac{1}{3}\Lambda r^2 \left(r^2+a^2 \right)
                +r^2 -2Mr + a^2,  \\                            \label{e2}
        \Delta_{\theta}& = & 1+ \frac{1}{3} \Lambda a^2 \cos^2 \theta, \\
                                                                \label{e3}
        I & = & 1+ \frac{1}{3} \Lambda a^2, \\                  \label{e4}
        \rho^2 & = & r^2 +a^2 \cos^2 \theta.                    \label{e5}
\eea
The parameters of the spacetime are mass ($M$), specific angular
momentum ($a$), and \cc\ ($\Lambda$). It is convenient to introduce a
dimensionless cosmological parameter
\be                                                             \label{e6}
        y = \frac{1}{3} \Lambda M^2.
\ee
For simplicity, we put $M=1$ hereafter. Equivalently, also the coordinates
$t,\ r$, the line element $\d s$, and the parameter of the spacetime $a$ are
expressed in units of $M$ and become dimensionless.

For $y<0$ corresponding to the attractive cosmological constant, the line element 
(\ref{e1}) describes a Kerr--anti--de~Sitter geometry. Here we focus our attention on 
the case $y>0$ corresponding to the repulsive cosmological constant, when Eq.~(\ref{e1}) 
describes a Kerr--de~Sitter spacetime.

\begin{figure}
\epsfxsize=.9\hsize
\epsfbox{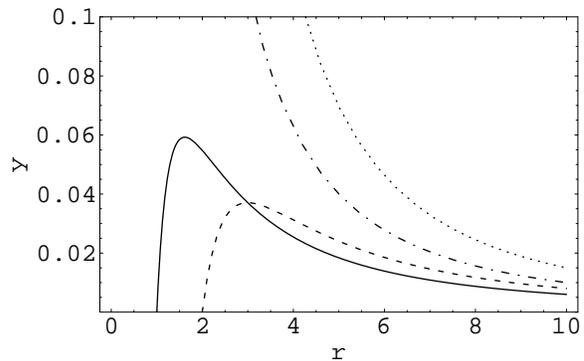}
\caption{Functions $y_{e(h)-}(r)$ (solid curve), $y_{z(h)}(r)$
(dashed curve), $y_{d(h)}(r)$ (dash-dotted curve), and
$y_{e(h)+}(r)$ (dotted curve) characterizing the properties of the function
$a^2_h(r;y)$ determining the horizons of \Kds spacetimes. The maximum
of $y_{e(h)-}(r)$ corresponds to the critical value of
the cosmological parameter $y_{c(KdS)}\doteq 0.05924$ for which the
function $a^2_h(r;y)$ has an inflex point (and no extrema). Any
\Kds spacetime with $y>y_{c(KdS)}$ is of naked-singularity type
containing only a cosmological horizon. The function
$y_{z(h)}(r)$ determines the horizons of the \Sds spacetimes and its
maximum $y_{c(SdS)}=1/27$ corresponds to the limiting value
for the existence of \Sds black holes.}
\label{f1}
\end{figure}

The event horizons of the spacetime are given by the pseudosingularities of
the line element (\ref{e1}), determined by the condition $\Delta_r =0$. The
loci of the event horizons are determined by the relation
\be                                                             \label{e7}
        a^2 = a^2_h(r;y) \equiv \frac{r^2 -2r -yr^4}{yr^2 -1}.
\ee
The asymptotic behavior of the function $a^2_h(r;y)$ is given by
$a^2_h(r\ra 0,y)\ra 0$, $a^2_h(r \ra \infty, y) \ra - \infty$. For $y=0$, the
function $a^2_h(r)= 2r-r^2$ determines loci of the horizons of Kerr black
holes. The divergent points of $a^2_h(r;y)$ are determined by
\be                                                             \label{e8}
        y = y_{d(h)}(r) \equiv \frac{1}{r^2},
\ee
its zero points are given by
\be                                                              \label{e9}
        y = y_{z(h)}(r) \equiv \frac{r-2}{r^3},
\ee
and its local extrema are determined by the relation
\be                                                             \label{e10}
        y = y_{\e(h)\pm}(r) \equiv \frac{2r+ 1 \pm \sqrt{8r+1}}
                {2r^3}.
\ee
The functions $y_{d(h)}(r)$, $y_{z(h)}(r)$, and $y_{\e(h)\pm}(r)$ are
illustrated in Fig.~\ref{f1}. 
The function $y_{\e(h)-}(r)$ has its maximum at
$r_{\rm crit} = (3+ 2\sqrt{3})/4$, where the value of the cosmological
parameter takes a critical value
\be                                                             \label{e11}
        y_{c(KdS)} = \frac{16}{(3+2\sqrt{3})^3} \doteq 0,05924;
\ee
for $y>y_{c(KdS)}$, only naked-singularity backgrounds exist for $a^2>0$.
A common point of the functions $y_{z(h)}(r)$ and $y_{e(h)-}(r)$ is located 
at $r=3$, where is the maximum of $y_{z(h)}(r)$ taking a critical value
\be                                                             \label{e12}
        y_{c(SdS)} = \frac{1}{27} \doteq 0.03704,
\ee
which is the limiting value for the existence of Schwarzschild--de~Sitter black
holes \cite{Stu-Hle:1999:PHYSR4:}. In Reissner--Nordstr\"{o}m--de~Sitter spacetimes,
the critical value of the cosmological parameter limiting the existence of
black-hole spacetimes is \cite{Stu-Hle:2002:ACTPS2:}
\be
        y_{c(RNdS)} = \frac{2}{27} \doteq 0.07407. 
                                                                \nonumber
\ee

If $y = y_{c(KdS)}$, the function $a^2_h(r;y)$ has an inflex point at
$r = r_{\rm crit}$, corresponding to a critical value of the rotation
parameter of Kerr--de~Sitter spacetimes:
\be                                                             \label{e13}
        a^2_{\rm crit} = \frac{3}{16} (3+ 2\sqrt{3}) \doteq 1,21202.
\ee
Kerr--de~Sitter black holes can exist for $a^2 < a^2_{\rm crit}$ only,
while Kerr--de~Sitter naked singularities can exist for both $a^2 <
a^2_{\rm crit}$ and $a^2 > a^2_{\rm crit}$.

\begin{figure}
\epsfxsize=.9\hsize
\epsfbox{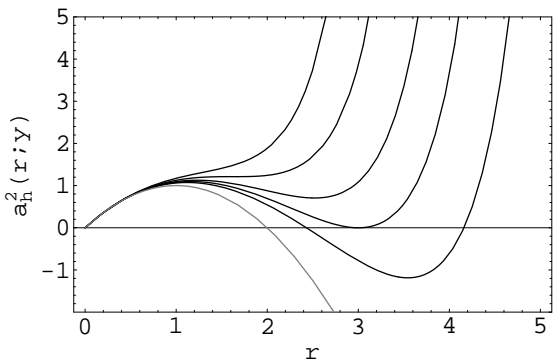}
\caption{Horizons of \Kds spacetimes. They are given for five typical
values of the cosmological parameter $y$ by the function $a^2_h(r;y)$. For 
$y>y_{c(KdS)}\doteq 0.05924\ (y=0.08)$ the function has no local extrema and only
naked-singularity spacetimes are allowed (the only horizon
is the cosmological horizon). For $y=y_{c(KdS)}$, the function has an inflex point 
where the black-hole and the cosmological horizons coincide. For $y_{c(SdS)}=1/27<y<y_{c(KdS)}\
(y=0.045)$ the function has two local extrema in positive values and the black-hole spacetimes 
exist for $a^2$ between those extrema. For $y=y_{c(SdS)}$ the local minimum resides on the axis $a^2=0$. 
The ciritical value $y_{c(SdS)}$ represents the limiting value of cosmological parameter for which
the \Sds black holes can exist; the \Kds black holes again exist for $a^2$ between those extrema. For
$0<y<y_{c(SdS)}\ (y=0.03)$ the local minimum resides in the nonphysical region
$a^2<0$ and the black holes  exist for $a^2$ up to the local maximum.
For completeness, we present the gray curve determining horizons of
the Kerr $(y=0)$ black holes. In all cases, the local extrema
correspond to the extreme black holes.} 
\label{f2}
\end{figure}

For $y > 0$, the function $y_{\e(h)-}(r)$ determines two local extrema of
$a^2_h(r;y)$ at $y< y_{c(KdS)}$, denoted as $a^2_{\max (h)}(r_1,y)$,
$a^2_{\min(h)}(r_2,y)$, with $r_1 < r_2$. If $y < y_{c(SdS)}$,
$a^2_{\min(h)}(r_2,y)< 0$, and the minimum is unphysical. The function $a^2_h(r)$
diverges at $r_d = 1/\sqrt{y}$, and it is discontinuous there. The function
$y_{\e(h)+}(r)$ determines a maximum of $a^2_h(r;y)$ at a negative value of
$a^2$ which is, therefore, physically irrelevant [see Fig.~\ref{f2} giving
typical behavior of $a^2_h(r;y)$].

\begin{figure}
\epsfxsize=.9\hsize
\epsfbox{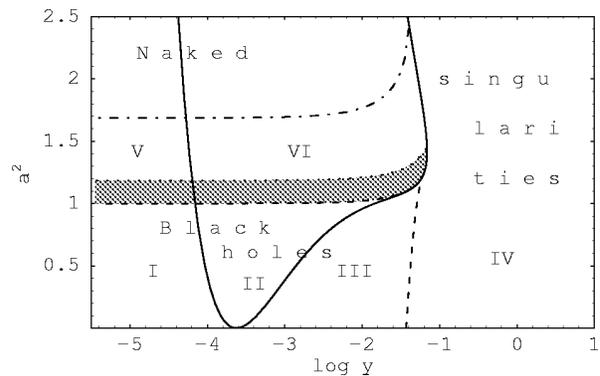}
\caption{Classification of Kerr--de~Sitter spacetimes. The space of 
parameters $a^2$ and $y$ is separated into six regions. Dashed curves separate 
regions of black holes and naked singularities. Solid curves divide the parametric 
space into spacetimes differing by properties of the stable circular orbits relevant 
for Keplerian accretion disks. For large values of $a^2$ both the solid lines tend to
the $a^2$ axis. ({\bf I}) Black-hole
spacetimes with both corotating and counterrotating stable or
bound circular orbits, ({\bf II}) black-hole spacetimes with no
counterrotating stable or bound circular orbits, ({\bf III}) black-hole
spacetimes with no corotating and counterrotating stable or bound
circular orbits, ({\bf IV}) naked-singularity spacetimes with no
corotating and counterrotating stable or bound circular orbits,
({\bf V}) naked-singularity spacetimes with both corotating and
counterrotating stable or bound circular orbits, and ({\bf VI})
naked-singularity spacetimes with no counterrotating stable or
bound circular orbits of the minus family. The dash-dotted curve
defines the subregion of naked-singularity spacetimes, where the 
plus-family circular orbits could be stable and counterrotating (from the point
of view of a locally nonrotating observer); shaded is the subregion
allowing stable circular orbits with $E_{+}<0$.}
\label{f3}
\end{figure}

If $0< y< y_{c(SdS)}$, black-hole spacetimes exist for $a^2 \leq
a^2_{\max(h)}(y)$, and naked-singularity spacetimes exist for $a^2
> a^2_{\max(h)}(y)$. If $y_{c(SdS)} <y \leq y_{c(KdS)}$, black-hole
spacetimes exist for $a^2_{\min(h)}(y) \leq a^2 \leq a^2_{\max(h)}(y)$, while
naked-singularity spacetimes exist for $a^2 <a^2_{\min(h)}(y)$ and $a^2 >
a^2_{\max(h)}(y)$. The functions $a^2_{\min(h)}(y)$, $a^2_{\max(h)}(y)$ are
implicitly given by Eqs.~(\ref{e7}) and (\ref{e10}); the separation of
Kerr--de~Sitter black-hole and naked-singularity spacetimes in the parameter 
space $y$--$a^2$ is shown in Fig.~\ref{f3}. 
In black-hole spacetimes, there
are two black-hole horizons and the cosmological horizon, with 
$r_{h-} < r_{h+} < r_{c}$. In naked-singularity 
spacetimes, there is the cosmological horizon $r_{c}$ only.

The extreme cases, when two (or all three) horizons coalesce, were
discussed in detail for the case of \RNds spacetimes \cite{Bri-Hay:1994:CLAQG:,%
                                                           Hay-Nak:1994:PHYSR4:}.
In \Kds spacetimes, the situation is analogical. If $r_{h-}=r_{h+} <
r_{c}$, the extreme black-hole case occurs; if $r_{h-} < r_{h+} =
r_{c}$, the marginal naked-singularity case occurs; if $r_{h-} =
r_{h+} = r_{c}$, the ``ultra-extreme'' case occurs which corresponds
to the naked-singularity case.  

\section{Equatorial motion}
\label{Eqmo}

In order to understand basic properties of thin accretion disks in the field of
rotating black holes or naked singularities, it is necessary to study equatorial
geodetical motion, especially circular motion, of test particles, as it can be 
shown that due to the dragging of the inertial frames any tilted disk has to be 
driven to the equatorial plane of the rotating spacetimes \cite{Bar-Pet:1975:ASTRJ2L:}.

\subsection{Carter equations}

The motion of a test particle with rest mass $m$ is given by the geodesic
equations. In a separated and integrated form,  the equations were obtained
by Carter \cite{Car:1973:BlaHol:}. For the motion restricted to the equatorial 
plane ($\d\theta/\d \lambda = 0$, $\theta = \pi/2$) the Carter equations take the
form
\bea
        r^2 \oder{r}{\lambda} & = & \pm R^{1/2} (r), \\
                                                                \label{e14}
        r^2 \oder{\phi}{\lambda} & = & - IP_{\theta}+
                \frac{a I P_r}{\Delta_r}, \\
                                                                \label{e15}
        r^2 \oder{t}{\lambda} & = & -a I P_{\theta} +
                \frac{(r^2 +a^2) I P_r}{\Delta_r},
                                                                \label{e16}
\eea
where
\bea
        R(r) & = & P^2_r -\Delta_r \left(m^2r^2 +K \right), \\
                                                                \label{e17}
        P_r & = & I {\cal{E}} \left(r^2 +a^2 \right) - I a \Phi, \\
                                                                \label{e18}
        P_{\theta} & = & I (a {\cal{E}} - \Phi), \\
                                                                \label{e19}
        K & = & I^2 (a {\cal{E}} - \Phi)^2.
                                                                \label{e20}
\eea
The proper time of the particle, $\tau$, is related to the affine parameter
$\lambda$ by $\tau = m \lambda$. The constants of the motion are energy
($\cal{E}$), related to the stationarity of the geometry; axial
angular momentum ($\Phi$), related to the axial symmetry of the
geometry; ``total'' angular momentum ($K$), related to the hidden
symmetry of the geometry. For the equatorial motion, $K$ is restricted through 
Eq.~(\ref{e20}) following from the conditions on the latitudinal motion 
\cite{Stu:1983:BULAI:}. Notice that $\cal{E}$ and $\Phi$ cannot be interpreted as 
energy and axial angular momentum at infinity, since the spacetime is not 
asymptotically flat.

\subsection{Effective potential}         

The equatorial motion is governed by the constants of motion
${\cal{E}},\ \Phi$. Its properties can
be conveniently determined by an ``effective potential'' given by the
condition $R(r) = 0$ for turning points of the radial motion. It is useful
to define specific energy and specific angular momentum by the relations
\be                                                     \label{e21}
        E \equiv \frac{I {\cal{E}}}{m}, \quad \ L \equiv \frac{I \Phi}{m}.
\ee
Solving the equation
\be                                                     \label{e22}
        R(r) \equiv \left[E\left(r^2 +a^2 \right)- a L \right]^2 -
                \Delta_r \left[r^2 +(aE -L)^2 \right] = 0,
\ee
we find the effective potential in the form
\begin{widetext}
\be                                                     \label{e23}
        E_{(\pm)}(r;L,a,y) \equiv \frac{a\left[yr\left(r^2+a^2 \right)+2
                \right]L \pm \Delta^{1/2}_r
                \left\{r^2 L^2 +r\left[\left(1+ya^2\right)r\left(r^2+a^2
                \right)+ 2a^2 \right] \right\}^{1/2}}
                {\left[\left(1+y a^2 \right)r\left(r^2 +a^2 \right)
                + 2a^2 \right]}.
\ee
\end{widetext}
In the stationary regions ($\Delta_r \geq 0$), the motion is allowed where
\be                                                            \label{e24}
        E \geq E_{(+)}(r;L,a,y)
\ee
or
\be                                                            \label{e25}
        E \leq E_{(-)}(r;L,a,y).
\ee
Conditions $E = E_{(+)}(r,L,a,y)$ [or $E = E_{(-)}(r;L,a,y)$] give the turning
points of the radial motion; at the dynamic regions ($\Delta_r < 0$), the
turning points are not allowed. In the region between the outer black-hole
horizon and the cosmological horizon, the motion of particles in the 
positive-root states---i.e., particles with positive energy as measured by 
local observers---being future directed ($\d t/\d \lambda > 0$) and having a 
direct ``classical" physical meaning, is determined by the effective potential 
$E_{(+)}(r;L,a,y)$. The character of the motion in the whole Kerr--de~Sitter 
background and the relevance of the effective potential $E_{(-)}(r;L,a,y)$,
determining the motion of particles in the negative-root states between the 
black-hole and cosmological horizons, is qualitatively the same as discussed 
in \cite{Bic-Stu-Bal:1989:BULAI:}. In the following we restrict our attention to the 
positive-root states determined by the effective potential $E_{(+)}(r;L,a,y)$.
Trajectories of the equatorial motion are then determined by the equation
\be                                                     \label{e26}
        \oder{\phi}{r} = \pm \frac{r}{\Delta_r}
                \frac{aE \left(r^2 +a^2 -\Delta_r \right) +
                \left(\Delta_r -a^2 \right) L}
                {\sqrt{\left[E \left(r^2+a^2 \right)- aL \right]^2 -
                \Delta_r \left[r^2 +\left(aE -L \right)^2 \right]}}.
\ee

Nevertheless, it is convenient to redefine the axial angular momentum by the relation
\be                                                     \label{e27}
        X \equiv L - aE;
\ee
for an analogous redefinition in the case of equatorial
photon motion see \cite{Stu-Hle:2000:CLAQG:}. With the constant of motion, $X$,
instead of $L$, the effective potential takes the simple form
\be                                                     \label{e28}
        E_{(+)}(r;X,a,y) \equiv \frac{1}{r^2}
                \left[aX +\Delta^{1/2}_r \left(r^2 +X^2
                \right)^{1/2}\right]
\ee
and the equation of trajectories (\ref{e26}) transforms to the form
\be                                                     \label{e29}
        \oder{\phi}{r} = \pm \frac{r}{\Delta_r}
                \frac{(r+2)X +aEr}{\sqrt{\left(r^2E- aX \right)^2 -
                \Delta_r \left(r^2 +X^2 \right)}}.
\ee

\section{Equatorial circular orbits}
\label{Eco}

\begin{figure*}
\epsfxsize=.7 \hsize
\epsfbox{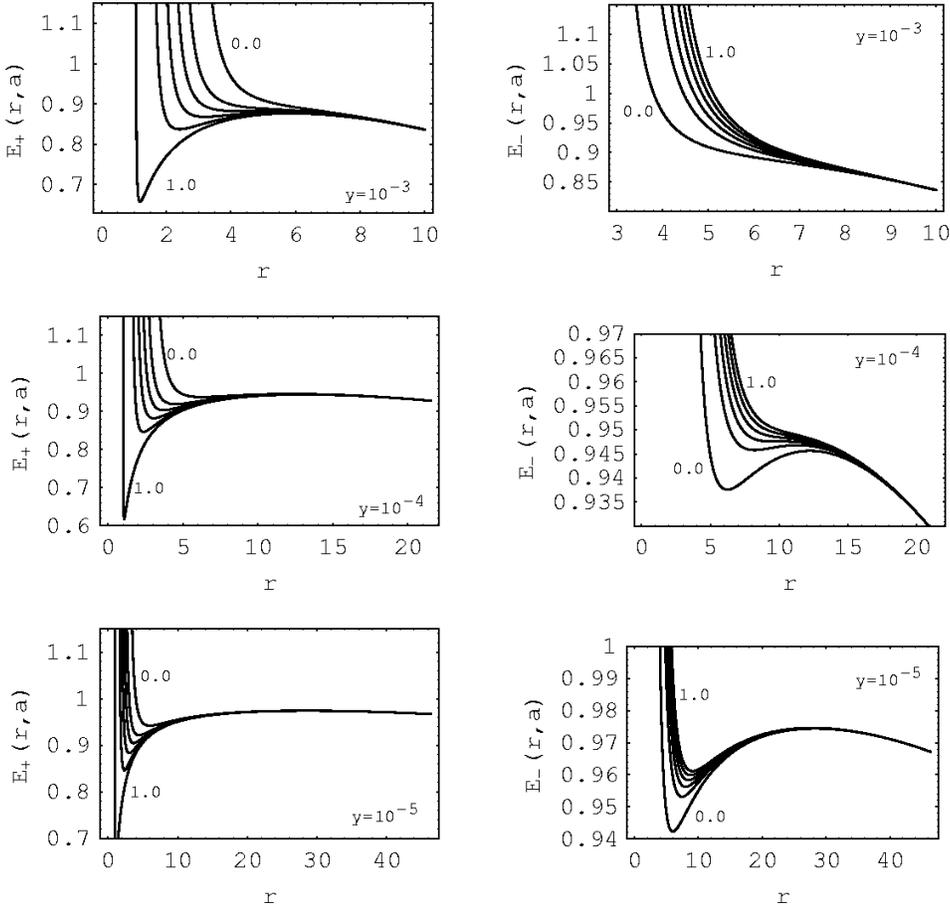}
\caption{Specific energy of the equatorial circular orbits in
Kerr--de~Sitter black-hole spacetimes. The spacetimes are specified
by the cosmological parameter $y$ and the rotational parameter $a$
($a^{2}$ varies from 0.0 to 1.0 in steps of 0.2). The left column
corresponds to the plus-family orbits; the right column corresponds
to the minus-family orbits. The local extrema of the curves correspond
to the marginally stable orbits, the rising parts correspond to stable orbits, and the
descending parts correspond to unstable ones. The behavior of the
curves for the spacetimes with $y<10^{-5}$ is similar to the case of
$y=10^{-5}$.}
\label{f4}
\end{figure*}

\begin{figure*}
\epsfxsize=0.7 \hsize
\epsfbox{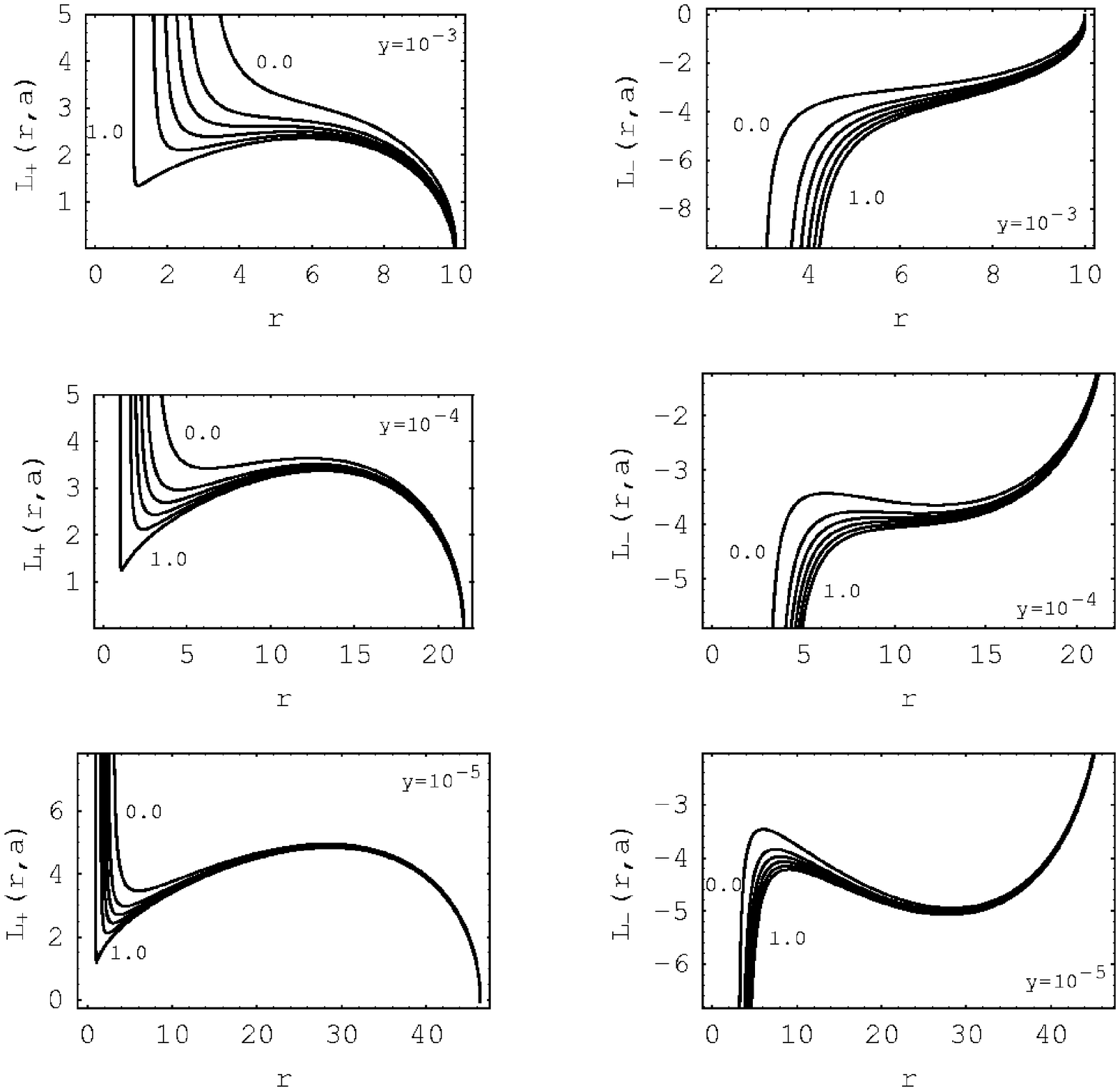}
\caption{Specific angular momentum of the equatorial circular orbits
in Kerr--de~Sitter black-hole spacetimes. The spacetimes are
specified by the cosmological parameter $y$ and the rotational parameter
$a$ ($a^{2}$ varies from 0.0 to 1.0 in steps of 0.2). The left
column corresponds to the plus-family orbits; the right column
corresponds to the minus-family orbits. The local extrema of the curves correspond
to the marginally stable orbits, the rising parts of $L_{+}$ and the descending
parts of $L_{-}$ correspond to the stable orbits, the descending parts of
$L_{+}$, and the rising parts of $L_{-}$ correspond to the unstable ones.
The behavior of the curves for the spacetimes with $y<10^{-5}$ is similar
to the case of $y=10^{-5}$.}
\label{f5}
\end{figure*}

\begin{figure*}
\epsfxsize=0.8 \hsize
\epsfbox{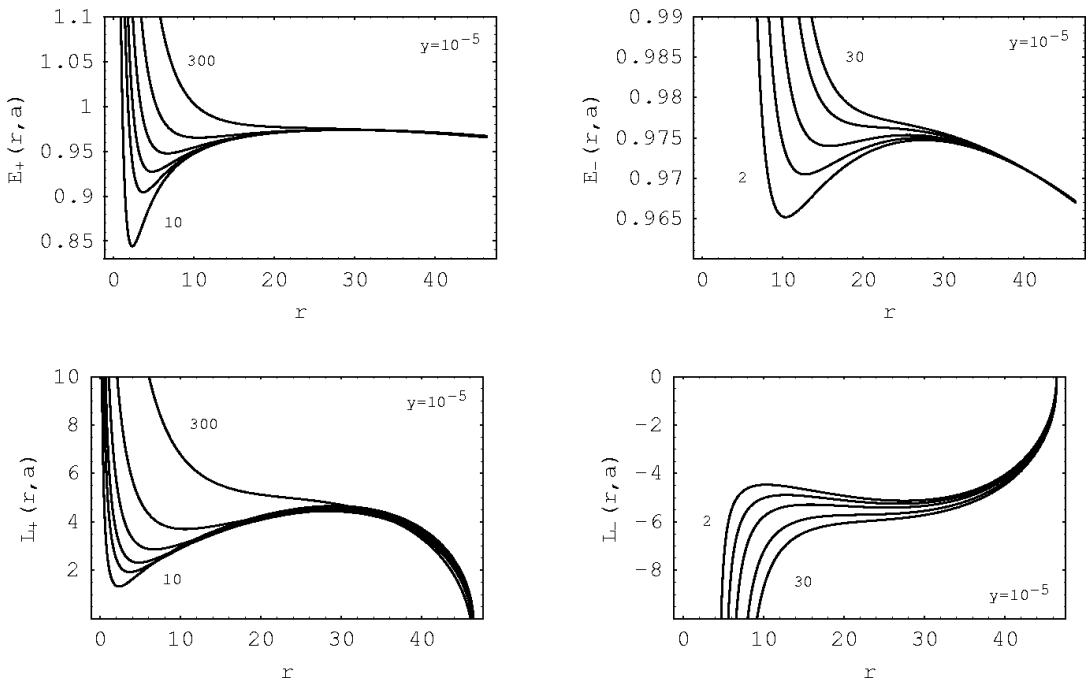}
\caption{Specific energy and specific angular momentum of the equatorial
circular orbits in Kerr--de~Sitter naked-singularity spacetimes.
The plus-family curves are plotted for the rotational parameter
$a^{2}=10, 20, 30, 50, 100, 300$; the minus-family curves are plotted
for $a^{2}=2, 5, 10, 20, 30$. The meaning of particular
parts of the curves is the same as in black-hole spacetimes.}
\label{f6}
\end{figure*}

The equatorial circular orbits can most easily be determined by
solving simultaneously the equations
\be                                                     \label{e30}
        R(r) = r^4 E^2 -2ar^2 EX +\left(a^2 -\Delta_r \right)X^2 -r^2
                \Delta_r = 0,
\ee
\be                                                     \label{e31}
        \oder{R}{r} = 4r^3 E^2 -4ar EX - \Delta_r' X^2
-\Delta_r'r^2
                - 2r \Delta_r = 0,
\ee
where $\Delta_r' \equiv \d\Delta_r /\d r$.
Combining Eqs.~(\ref{e30}) and (\ref{e31}), we arrive at a quadratic equation
\be                                                     \label{e32}
        A(r) \left(\frac{X}{E}\right)^2 + B(r)\left(\frac{X}{E}\right)
                + C(r) = 0,
\ee
with
\bea
        A(r) & = & 2 \Delta_r \left(a^2 -\Delta_r\right)+
                        a^2 \Delta'_r r,  \\            \label{e33}
        B(r) & = & - 2a \Delta'_r r^3, \\
                                                        \label{e34}
        C(r) & = & r^4 \left(\Delta'_r r - 2 \Delta_r \right).
                                                        \label{e35}
\eea
Its solution can be expressed in the relatively simple form
\bea                                                     
       \lefteqn{\left(\frac{X}{E}\right)_{\pm}(r;a,y)=}
                                                    \nonumber \\
            & & \frac{r^2 \left(r- a^2- yr^4 \right)}
                {ar\left[r- 1- yr\left(2r^2+ a^2 \right)\right]
                \pm \Delta_r \left[r\left(1- yr^3\right)\right]^{1/2}}.             
                                                    \nonumber \\
                                                        \label{e36}
\eea
Assuming now 
\be                                                     \label{e37}
        X_{+} = E_{+}\left(\frac{X}{E}\right)_{+},
                \ X_{-} = E_{-}\left(\frac{X}{E}\right)_{-},
\ee
substituting into Eq.~(\ref{e30}), and solving for the specific energy of the orbit,
we obtain
\be                                                     \label{e38}
        E_{\pm}(r;a,y) = \frac{1-\frac{2}{r}- \left(r^2+ a^2 \right)y
                \pm a \left(\frac{1}{r^3}- y\right)^{1/2}}
                {\left[1- \frac{3}{r}- a^2y\pm 2a \left(\frac{1}{r^3}- y
                \right)^{1/2} \right]^{1/2}}.
\ee
The related constant of motion, $X$, of the orbit is then given by the expression
\be                                                     \label{e39}
        X_{\pm}(r;a,y) = \frac{-a\pm r^2 \left(\frac{1}{r^3}- y
                \right)^{1/2}}
                {\left[1-\frac{3}{r}- a^2y\pm 2a\left(\frac{1}{r^3}
                -y \right)^{1/2}\right]^{1/2}},
\ee
while the specific angular momentum of the circular orbits is determined by
the relation
\bea                                                     
       \lefteqn{L_{\pm}(r;a,y) =} \nonumber \\
           & &  - \frac{2a +ar\left(r^2 +a^2 \right)y
                \mp r\left(r^2 +a^2\right)
                \left(\frac{1}{r^3} -y \right)^{1/2}}
                {r \left[1 -\frac{3}{r}- a^2y\pm 2a
                \left(\frac{1}{r^3} -y\right)^{1/2} \right]^{1/2}}. \nonumber \\
                                                         \label{e40}
\eea
Relations (\ref{e38})--(\ref{e40}) determine two families of the circular orbits. We call them
plus-family orbits and minus-family orbits according to the $\pm$ sign in relations
(\ref{e38})--(\ref{e40}). The typical behavior of the functions $E_{\pm}(r;a,y)$ and $L_{\pm}(r;a,y)$
giving the specific energy and specific angular momentum is illustrated in Figs.~\ref{f4} 
and \ref{f5}, respectively, for \Kds black-hole spacetimes with appropriately taken
parameters. 
Figure~\ref{f6} shows the typical behavior of these functions for some Kerr--de~Sitter
naked-singularity spacetimes.

In the limit of $y \ra 0$, relations (\ref{e38}) and (\ref{e40}) reduce to the expression 
given by Chandrasekhar (in units of $M$) \cite{Chan:1983:BlackHoles:} for circular orbits in the Kerr 
backgrounds:
\bea
        E_{\pm}(r;a) & = & \frac{1- \frac{2}{r}\pm
                \frac{a}{r^{3/2}}}
                {\left[1 -\frac{3}{r}\pm
                \frac{2a}{r^{3/2}}\right]^{1/2}},  \label{e41}
        \\
        L_{\pm}(r;a) & = & \pm r^{1/2}\frac{1+ \frac{a^2}{r^2}
                \mp \frac{2a}{r^{3/2}}}
                {\left[1- \frac{3}{r}\pm \frac{2a}{r^{3/2}}\right]^{1/2}}.
                                                        \label{e42}
\eea

In the limit of $a \ra 0$ we arrive at the formulas determining the specific 
energy and the specific angular momentum of circular orbits in the field of 
Schwarzschild--de~Sitter black holes \cite{Stu-Hle:1999:PHYSR4:}:
\bea
        E(r;y) & = & \frac{r -2 -yr^3}
                {\left[r(r-3) \right]^{1/2}}, \\
                                                        \label{e47}
        L(r;y) & = & \frac{r\left(1 -yr^3 \right)^{1/2}}
                {(r -3)^{1/2}};
                                                        \label{e48}
\eea
here, we do not give $L$ for the minus-family orbits as these are equivalent
to the plus-family orbits in spherically symmetric spacetimes.

The formulas for the specific energy and angular momentum of the equatorial circular 
orbits hold equally for both Kerr--de~Sitter ($y>0$) and Kerr--anti-de~Sitter ($y<0$)
spacetimes. Here, we shall concentrate our discussion on the circular motion in 
Kerr--de~Sitter spacetimes. We shall determine radii where the existence of circular 
orbits is allowed, the orientation of the circular motion relative to the locally nonrotating
frames, stability of the circular motion relative to radial perturbations. Finally, we 
shall introduce the notion of marginally bound orbits.

\subsection{Existence of circular orbits}

Inspecting expressions (\ref{e38}) and (\ref{e40}), we find two reality
conditions on the circular orbits. The first restriction on the existence of
circular orbits is given by the relation
\be                                                     \label{e49}
        y \leq y_s \equiv \frac{1}{r^3},
\ee
which introduces the notion of the ``static radius,'' given by the formula 
$r_s = y^{-1/3}$ independently of the rotational parameter $a$. It can be compared with 
the formally identical result in Schwarzschild--de~Sitter spacetimes
\cite{Stu-Hle:1999:PHYSR4:}. A ``free'' or ``geodetical'' observer on the static radius has 
only $U^t$ component of four-velocity nonzero. The position on the static radius is unstable 
relative to radial perturbations, as follows from the discussion on stability of the circular 
orbits performed below. 

The second restriction on the existence of circular orbits is given by the condition
\be                                                     \label{e50}
        1 -\frac{3}{r} -a^2y \pm 2a\left(\frac{1}{r^3}
                -y \right)^{1/2} \geq 0;
\ee
the equality determines the radii of photon circular orbits, where both $E \ra
\infty$ and $L \ra \pm \infty$.

\begin{figure*}
\epsfxsize=.775 \hsize
\epsfbox{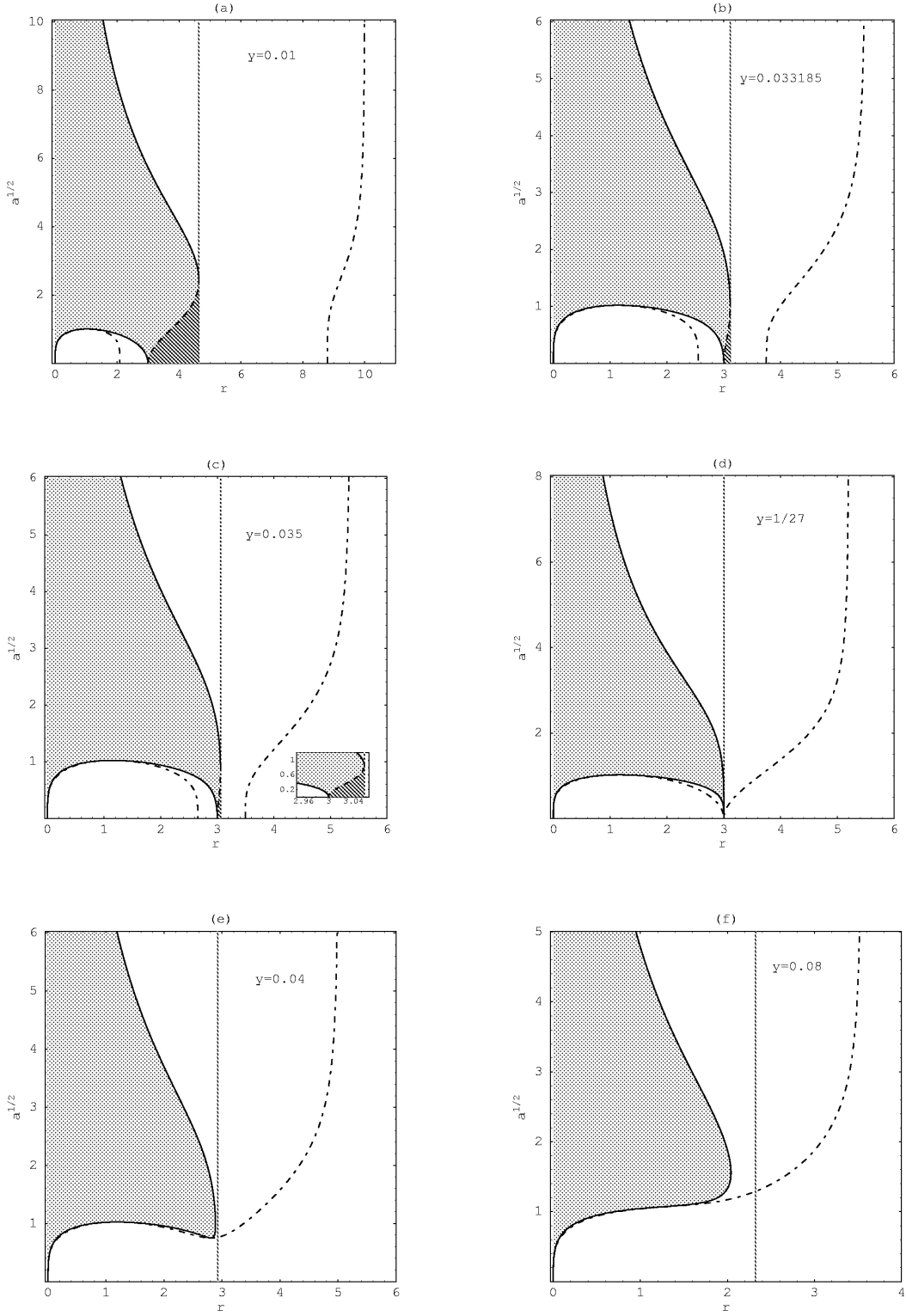}
\caption{Existence of the circular orbits. Locations of the horizons and the
photon circular orbits are given as functions of the rotational parameter for
typical values of the cosmological parameter of Kerr--de~Sitter spacetimes
and the regions of the existence of the circular orbits are shown.
Dash-dotted curves depict the horizons, solid curves depict the plus-family
photon circular orbits, dashed curves depict the minus-family photon circular 
orbits, and the gray line gives the static radius ($r_s = y^{-1/3}$). Notice 
that the static radius is relevant only in the spacetimes admitting the minus-family
circular orbits. The plus-family circular orbits exist in the whole shaded region,
while the minus-family circular orbits exist in the dark-gray regions only---i.e., in the 
spacetimes with $y < y_{c(SdS)}$. In the naked-singularity spacetimes only one
photon circular orbit exists and the plus-family orbits always approach the ring
singularity at $r=0$.}
\label{f7}
\end{figure*}

The photon circular orbits of the plus-family are given by the relation
\be                                                     \label{e51}
       a = a^{(+)}_{ph (1,2)}(r;y) \equiv \frac{\left(1 -yr^3\right)^{1/2}
                \pm \left(1 -3yr^2 \right)^{1/2}}
                {yr^{3/2}},
\ee
while for the minus-family orbits they are given by the relation
\be                                                     \label{e52}
      a = a^{(-)}_{ph (1,2)}(r;y) \equiv \frac{-\left(1 -yr^3 \right)^{1/2}
                \pm \left(1 -3yr^2 \right)^{1/2}}
                {yr^{3/2}}.
\ee
The photon circular orbits can be determined by a ``common'' formula related to $a^2$:
\bea                                                    
        \lefteqn{a^2 = a^2_{ph (1,2)}(r;y) \equiv} \nonumber \\
           & &  \frac{\left(1 -yr^3 \right) +
                \left(1 -3yr^2 \right)\pm 2 \sqrt{\left(1 -yr^3\right)
                \left(1 -3yr^2 \right)}} {y^2 r^3}, \nonumber \\
                                             \label{e53}
\eea
where the notation $a^2_{ph(1)}$ and $a^2_{ph(2)}$ is used for the $\pm$ parts of Eq.~(\ref{e53}) 
because these functions can define photon orbits for both plus- (minus-) family circular orbits. 
A detailed discussion of the equatorial photon motion is presented in \cite{Stu-Hle:2000:CLAQG:}, where more 
general, Kerr--Newman--(anti-)de~Sitter spacetimes are studied. In the Kerr--de~Sitter spacetimes
the situation is much simpler. Since
\begin{widetext}
\be                                                     \label{e54}
  \pder{a^2_{ph (1,2)}}{r} =
    \frac{\left[1 -yr^2(r+3) +3y^2r^5
                \right]^{1/2}\left(yr^2 -2 \right) \pm \left[-2
                +yr^2 (r+4) -y^2 r^5 \right]}
                {y^2 r^4 \left[1 -yr^2 (r+3) +3y^2r^5 \right]^{1/2}},
\ee
\end{widetext}
we find that the local extrema of $a^2_{ph (1,2)}(r;y)$ are located at radii
determined by the relation
\be                                                     \label{e55}
        y = y_{\e(ph)\pm}(r) \equiv \frac{2r +1 \pm \sqrt{8r +1}}
                {2r^3}
                = y_{\e(h)\pm}(r).
\ee
Therefore, $a^2_{ph (1,2)}(r;y)$ and $a^2_h(r;y)$ have common points at their local extrema. 
Nevertheless, in order to obtain directly limits on the existence of the plus- (minus-) family
circular orbits, it is convenient to consider the plus- (minus-) photon circular orbits determined
by relations (\ref{e51}) and (\ref{e52}), respectively, under the assumption $a \geq 0$.
We have to introduce a critical value of the rotational parameter corresponding to the situation
where $a^{(+)}_{ph (1)}(r;y) = a^{(-)}_{ph (1)}(r;y)$---i.e., where these functions reach the 
static radius $r_s = y^{-1/3}$:
\be
           a^2_{c(s)}(y) \equiv \frac{1 - 3y^{1/3}}{y}.             \label{s1}
\ee
Further, it is necessary to determine (by a numerical procedure) the related critical value of 
the cosmological parameter $y_{c(s)}$ such that for $y < y_{c(s)}$ the critical value $a^2_{c(s)}(y)$
corresponds to a naked-singularity spacetime. The numerical procedure implies
\be
                 y_{c(s)} \doteq 0.033185.
\ee
The results can be summarized in dependence on the cosmological parameter and are illustrated in
Fig.~\ref{f7}.
\begin{enumerate}
\item
$y \leq y_{c(s)}$   (Figs.~\ref{f7}a,\ref{f7}b)

In black-hole spacetimes there are three photon circular orbits. Their loci satisfy the conditions
\be
      r_{ph(1)} < r_{h(-)} < r_{h(+)} < r_{ph(2)} < r_{ph(3)} < r_{s}.     \label{s2}
\ee
The orbits $r_{ph(1)}$ and $r_{ph(2)}$ belong to the plus-family orbits, while
$r_{ph(3)}$ belongs to the minus-family orbits. 
We can conclude that in the black-hole backgrounds, the plus-family circular orbits 
are located at radii satisfying the relations
\be
   0 < r < r_{ph(1)} \qquad\mbox{and}\qquad r_{ph(2)} < r < r_{s},         \label{s3}
\ee
while the minus-family orbits are located at radii satisfying the relation
\be
       r_{ph(3)} < r < r_{s}.                                              \label{s4}
\ee

In the naked-singularity spacetimes, we have to distinguish two qualitatively different cases.

If $a^2 \leq a^2_{c(s)}(y)$, there is one photon circular orbit belonging to the minus-family 
orbits. In such spacetimes, the plus-family orbits are located in the region
\be
      0 < r < r_{s}.                                          \label{s5}
\ee

If $a^2 > a^2_{c(s)}(y)$, the situation changes dramatically as the minus-family orbits (and the
notion of the static radius) cease to exist. There is only one plus-family photon circular orbit.
Therefore, the plus-family circular orbits are located in the region
\be
      0 < r < r_{ph}.                                \label{s6}
\ee
\item
$y_{c(s)} < y \leq y_{c(SdS)}$       (Figs.~\ref{f7}c,\ref{f7}d)

Now, we have to distinguish two cases in black-hole spacetimes.

If $a^2 \leq a^2_{c(s)}(y)$, the loci of the photon circular orbits are again related by relation
(\ref{s2}), and the limits on the existence of plus-family and minus-family circular orbits 
are the same as in the case of $y < y_{c(s)}$---see relations (\ref{s3}) and (\ref{s4}), 
respectively.

If $a^2 > a^2_{c(s)}(y)$, black-hole spacetimes admit only the plus-family circular orbits and all 
of the three photon circular orbits limit them by the relation
\be
   0 < r < r_{ph(1)} \quad\mbox{and}\quad r_{ph(2)} < r < r_{ph(3)}.         \label{s7}
\ee

In naked-singularity spacetimes, only one plus-family photon circular
orbit exists and the plus-family circular orbits are limited by relation (\ref{s6}).
\item
$y_{c(SdS)} < y \leq y_{c(KdS)}$       (Fig.~\ref{f7}e)

Only the plus-family circular orbits exist that are limited by three photon circular orbits through
relation (\ref{s7}) in black-hole spacetimes and by one photon circular orbit through relation
(\ref{s6}) in naked-singularity spacetimes.
\item
$y_{c(KdS)} < y$       (Fig.~\ref{f7}f)

Naked-singularity spacetimes exist for any $a > 0$. The spacetimes admit only the plus-family 
circular orbits limited by one photon circular orbit through relation (\ref{s6}).
\end{enumerate}

\subsection{Orientation of the circular orbits}\label{orientation}

The behavior of the circular orbits in the field of Kerr black holes ($y=0$) 
suggests that the plus-family orbits correspond to the corotating orbits, 
while the minus-family circular orbits correspond to the counterrotating ones. 
However, this statement is not generally correct even in some of the 
Kerr naked-singularity spacetimes---namely, in the spacetimes with the rotational parameter low enough, 
where counterrotating plus-family orbits could exist nearby the ring singularity 
\cite{Stu:1980:BULAI:}. In Kerr--de~Sitter spacetimes, the situation is even 
more complicated and we cannot identify the plus-family circular orbits with purely
corotating orbits even in black-hole spacetimes. Moreover, in rotating 
spacetimes with a nonzero cosmological constant it is not possible to define the 
corotating (counterrotating) orbits in relation to stationary observers at infinity,
as can be done in Kerr spacetimes, since these spacetimes are not asymptotically flat.

The natural way of defining the orientation of the circular orbits in Kerr--de~Sitter 
spacetimes is to use the point of view of locally nonrotating frames that is used in
asymptotically flat Kerr spacetimes too. The tetrad of one-forms corresponding
to these frames in the Kerr--de~Sitter backgrounds is given by \cite{Stu-Hle:2000:CLAQG:}
\begin{eqnarray}
\omega^{(t)}\equiv\left (\frac{\Delta_{r}\Delta_{\theta}\varrho^{2}}{I^{2}A}\right )^{1/2}{\rm d}t,\\
\omega^{(\phi)}\equiv\left (\frac{A\sin^{2}\theta}{I^{2}\varrho^{2}}\right )^{1/2}({\rm d}\phi-\Omega {\rm d}t),\\
\omega^{(r)}\equiv\left (\frac{\varrho^{2}}{\Delta_{r}}\right )^{1/2}{\rm d}r,\\
\omega^{(\theta)}\equiv\left (\frac{\varrho^{2}}{\Delta_{\theta}}\right )^{1/2}{\rm d}\theta,
\end{eqnarray}
where
\begin{equation}
A\equiv (r^{2}+a^{2})^{2}-a^{2}\Delta_{r},
\end{equation}
\begin{equation}
   \Delta_{\theta} \equiv 1 + y a^2 cos^2\theta ,
\end{equation}
and the angular velocity of the locally nonrotating frames,
\begin{equation}
\Omega\equiv\frac{{\rm d}\phi}{{\rm d}t}=\frac{a}{A}\left [-\Delta_{r}+(r^{2}+a^{2})\Delta_{\theta}\right ].
\end{equation}
Note that $\Delta_{\theta}=1$ in the equatorial plane.

Locally measured components of the four-momentum are given by the projection
of a particle's four-momentum  onto the tetrad:
\begin{equation}
p^{(\alpha)}=p^{\mu}\omega^{(\alpha)}_{\mu},
\end{equation}
where
\begin{equation}
p^{\mu}=m\frac{{\rm d}x^{\mu}}{{\rm d}\tau}\equiv m\dot{x}^{\mu}=\frac{{\rm d}x^{\mu}}{{\rm d}\lambda}
\end{equation}
are the coordinate components of particle's four-momentum, the affine parameter $\lambda=\tau/m$,
$m$ denotes the rest mass of the particle, and $\tau$ is its proper time.

In the equatorial plane, $\theta=\pi /2$, the azimuthal component of the four-momentum measured in the
locally nonrotating frames is given by the relation
\begin{equation}
p^{(\phi)}=\frac{mA^{1/2}}{Ir}\left (\dot{\phi}-\Omega\dot{t}\right ),
\end{equation}
where the temporal and azimuthal components of the four-momentum, determined by the geodesic equations,
can be expressed in the form containing the specific constants of motion $E,\ X$:
\begin{eqnarray}
\dot{t} &=& \frac{I}{r^{2}}\left [aX+\frac{(r^{2}+a^{2})(r^{2}E-aX)}{\Delta_{r}}\right ],\\
\dot{\phi} &=& \frac{I}{r^{2}}\left [X+\frac{a}{\Delta_{r}}(r^{2}E-aX)\right ]\label{phidot}.
\end{eqnarray}
A simple calculation reveals
\begin{equation}
p^{(\phi)}=\frac{mr}{A^{1/2}}(aE+X)
\end{equation}
and using Eq.~(\ref{e27}) we obtain intuitively anticipated relation
\begin{equation}
p^{(\phi)}=\frac{mr}{A^{1/2}}L.
\end{equation}
We can see that the sign of the azimuthal component of the four-momentum measured in the locally
nonrotating frames is given by the sign of the specific angular momentum of a particle on the orbit of 
interest. Therefore, the circular orbits with $p^{(\phi)}>0$ $(L>0)$ we call corotating, and the circular
orbits with $p^{(\phi)}<0$ $(L<0)$ we call counterrotating, in agreement with the approach used in 
the case of asymptotically flat Kerr spacetimes.

\subsection{Stability of the circular orbits}

\begin{figure}
\epsfxsize=0.9 \hsize
\epsfbox{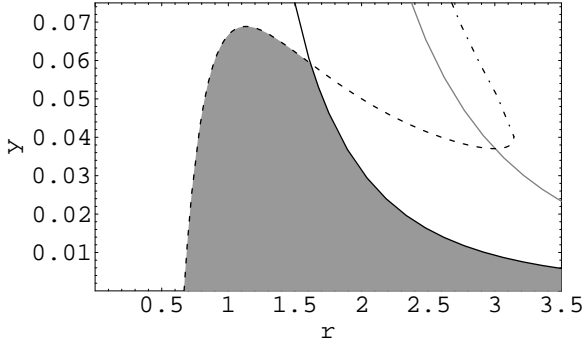}
\caption{Reality conditions for the existence of the stable circular orbits. Black and gray 
solid curves correspond to the functions $y_{ms}(r)$ and $y_s(r)$, respectively; dash-dotted 
and dashed curves correspond to the functions $y_{ms+}(r)$ and $y_{ms-}(r)$, respectively.
Stable orbits can exist only in the shaded region, where the local maximum corresponds to
the critical value of the cosmological parameter $y_{\rm crit(ms+)}\doteq 0.06886$.}
\label{f8}
\end{figure}
The loci of the stable circular orbits are given by the condition
\be                                                     
        \frac{\d^2 R}{\d r^2} \geq 0,
                                     \label{e56}
\ee
which has to be satisfied simultaneously with the conditions $R(r) =0$ and
$\d R/\d r = 0$ determining the specific energy and the specific angular momentum of
the circular orbits. Using relations (\ref{e38}) and (\ref{e39}), we find that the radii 
of the stable orbits of both families are restricted by the condition
\bea                                                     
        \lefteqn{r \left[6 -r +r^3 (4r -15)y \right] \mp} \nonumber \\
           & &   \mp 8a \left[r\left(1 -yr^3 \right)^3\right]^{1/2} + 
                 a^2 \left[3 +r^2y\left(1 -4yr^3 \right)\right] \geq 0. 
                                                          \nonumber \\
                                                \label{e57}
\eea

The marginally stable orbits of both families can be described together by the relation
\begin{widetext}
\bea
  a^2=a^2_{ms(1,2)}(r;y) & \equiv &
    \left[3 +r^2y \left(1-4yr^3\right)\right]^{-2}r
      \left\{\left[r -6 -r^3 (4r -15)y \right]\right. \nonumber \\
    & & \times
      \left[3 +r^2y \left(1 -4yr^3 \right)\right]
      +32 \left(1 -yr^3 \right)^3 \pm 8 \left(1-yr^3\right)^{3/2}
      \left(1 -4yr^3 \right)^{1/2} \nonumber \\
    & & \times \left.\left\{r \left[3 -ry \left(6 +10r -15yr^3\right) \right]
      -2 \right \}^{1/2}\right \}.    \label{e58}
\eea
\end{widetext}
\begin{figure*}
\epsfxsize=.95 \hsize
\epsfbox{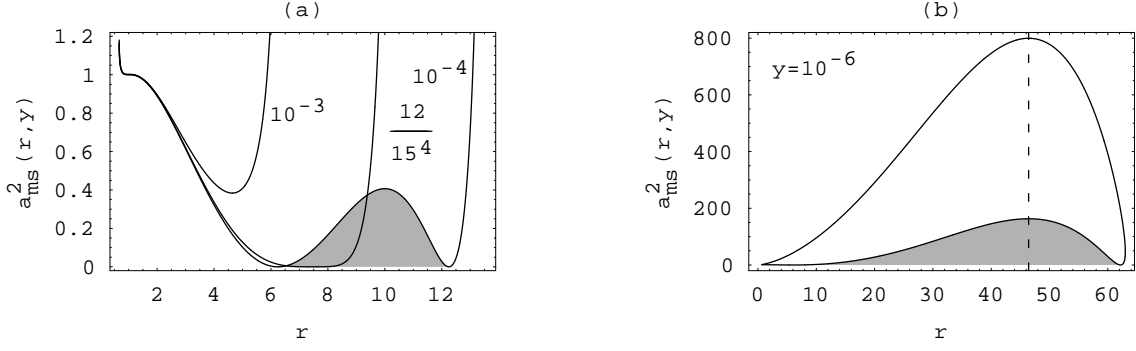}
\caption{Marginally stable circular orbits in Kerr--de~Sitter spacetimes. The relevant 
functions are given for some typical values of the cosmological parameter $y$. {\bf (a)}
The black-hole region of Kerr--de~Sitter spacetimes. For
$y<12/15^{4}$ there exist spacetimes containing four
marginally stable (ms) orbits. For a given spacetime, the innermost and the outermost ms orbits
belong to the plus-family orbits; the two orbits in between belong to the minus-family orbits. 
{\bf (b)} In the naked-singularity region there exist spacetimes with no stable
orbits for a fixed value of $y$ [spacetimes with $a^{2}$ greater than
the global maximum of function $a^{2}_{ms}(r;y)$ for a given $y$].
Stable counterrotating (minus-family) orbits exist only in shaded
regions of the presented spacetimes. But some naked-singularity spacetimes contain 
counterrotating plus-family orbits; for more details, see the text. The dashed line corresponds to
the radius $(10y)^{-1/3}$ where both maxima of $a^2_{ms}(r;y)$
are located.}
\label{f9}
\end{figure*}
The ($\pm$) sign in Eq.~(\ref{e58}) is not directly related to the plus-family and
minus-family orbits. The function $a^2_{ms(1)}$, corresponding to the $+$ sign in Eq.
(\ref{e58}), determines marginally stable orbits of the plus-family orbits, while the function
$a^2_{ms(2)}$, corresponding to the $-$ sign in Eq.~(\ref{e58}), is relevant for both
the plus-family and minus-family orbits.
The reality conditions for the functions $a^2_{ms(1,2)}(r;y)$ are directly given by 
Eq.~(\ref{e58}). The standard condition $y \leq y_s(r) \equiv 1/r^3$ is guaranteed
by the first relevant condition
\be                                                     \label{e59}
        y \leq y_{ms}(r) \equiv \frac{1}{4r^3}.
\ee
The other two conditions can be given in the form
\be                                                     \label{e60}
        y \leq y_{ms-}(r)\qquad \mbox{or}\qquad y \geq y_{ms+}(r),
\ee
where the functions $y_{ms(\pm)}(r)$ are given by the relation
\be                                                     \label{e61}
        y_{ms\pm}(r) = \frac{3 +5r\pm \left(60r -20r^2 +9
                \right)^{1/2}} {15r^3}.
\ee
The behavior of the functions $y_{s}(r)$, $y_{ms}(r)$, and $y_{ms{\pm}}(r)$ 
is illustrated in Fig.~\ref{f8}. 
The function $y_{ms(+)}(r)$ is irrelevant;
the relevant function $y_{ms(-)}(r)$ intersects the function $y_{s}(r)$ at 
$r=3$, where $y = y_{c(SdS)}=1/27$, and the function $y_{ms}(r)$ at $r = (3 + 2\sqrt{3})/4$,
where $y=y_{i}=16/(3+2\sqrt{3})^3$. The critical value of the cosmological
parameter for the existence of the stable (plus-family) orbits, corresponding to 
the local maximum of $y_{ms(-)}(r)$, is given by
\be
          y_{crit(ms+)} = \frac{100}{(5 + 2\sqrt{10})^{3}}\doteq 0.06886.
                                                        \label{ee1}
\ee
The related critical value of the rotational parameter is
\be
          a^2_{crit(ms+)} = \frac{955 + 424\sqrt{10}}{1620} \doteq 1.41716.
                                                        \label{ee2}
\ee
The plus-family stable circular orbits are allowed for $y < y_{ms}(r)$, if $y < y_i$,
and for $y < y_{ms(-)}(r)$, if $y_i < y < y_{crit(ms+)}$. 

The condition determining the local extrema of $a^2_{ms(1,2)}(r;y)$, 
\be                                                     \label{e62}
        \frac{\p a^2_{ms(1,2)}(r;y)}{\p r} = 0,
\ee
implies very complicated relations; however, they lead to one simple relevant
relation
\be                                                     \label{e64}
        y = y_{\e (ms)}(r) \equiv \frac{1}{10r^3}
\ee
determining important local extrema of both $a^{2}_{ms(1,2)}(r;y)$
simultaneously, both located on the radius 
\be
        r=r_{e(ms)}(y)\equiv\frac{1}{(10y)^{1/3}}.       
\ee

The critical value of the cosmological parameter for the existence of the minus-family
stable circular orbits, determined by the condition $a^{2}_{ms(2)}(r_{e(ms)};y)=0$, is given by
\be
          y_{crit(ms-)} = \frac{12}{15^{4}}.             \label{ee3}
\ee
It coincides with the limit on the existence of the stable circular orbits in 
Schwarzschild--de~Sitter spacetimes \cite{Stu-Hle:1999:PHYSR4:}.

Properties of the functions $a^{2}_{ms(1,2)}(r;y)$ can be summarized in the 
following way.
\begin{enumerate}
\item
$y > y_{crit(ms+)}$. No stable circular orbits are allowed for any value of the rotational 
parameter.
\item
$y_{crit(ms+)} > y > y_{crit(ms-)}$. At $r=r_{e(ms)}$, the function $a^{2}_{ms(1)}(r;y)$ has a local maximum
($a^{2}_{ms(max)}$), and the function $a^{2}_{ms(2)}(r;y)$ has a local
minimum ($a^{2}_{ms(min)}$). For $a^{2}_{ms(min)} < a^{2} < a^{2}_{ms(max)}$,
the equation $a^{2} = a^{2}_{ms(1,2)}(r;y)$ determines two marginally stable
plus-family circular orbits (an inner one and an outer one). For 
$0 < a^{2} < a^{2}_{ms(min)}$ and $a^{2} > a^{2}_{ms(max)}$, no stable circular
orbits are allowed.
\item
$y < y_{crit(ms-)}$. There are two zero points of the function $a^{2}_{ms(2)}(r;y)$ corresponding to
its local minima, while it has a local maximum $a^{2}_{ms(max2)}$ at $r=r_{e(ms)}$,
where the maximum of the function $a^{2}_{ms(1)}(r;y)$ is located too.
For $a^{2} > a^{2}_{ms(max)}$, there is no stable circular orbit. 
For $a^{2}_{ms(max2)} < a^{2} < a^{2}_{ms(max)}$, there are two marginally
stable plus-family circular orbits. For $a^{2} < a^{2}_{ms(max2)}$, there are
four marginally stable orbits. The innermost and the outermost orbits belong to the
plus-family orbits; the two orbits in between belong to the minus-family orbits. 
\end{enumerate}

The functions $a^2_{ms(1,2)}$ are illustrated for typical values of the cosmological
parameter in Fig.~\ref{f9}.
In the parameter space $y$-$a^{2}$, separation of
Kerr--de~Sitter spacetimes according to the existence of stable circular orbits,
determined by the functions $a^{2}_{ms(1,2)}(r;y)$ and $y_{e(ms)}(r)$, is given 
in Fig.~\ref{f3}.

\subsection{Marginally bound circular orbits}

\begin{figure}
\epsfxsize=.99\hsize
\epsfbox{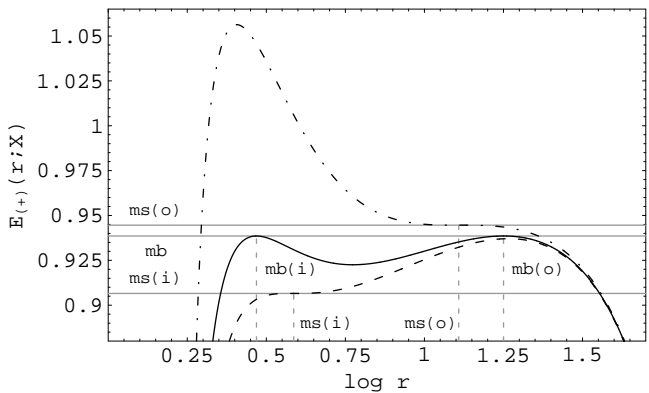}
\caption{Effective potential of the equatorial radial motion of test particles in an appropriatelly 
chosen \Kds black-hole spacetime ($y=10^{-4},\ a^2=0.36$) 
allowing stable circular orbits for corotating particles. Marginally bound (mb) orbits are given 
by the solid curve corresponding to the angular momentum parameter $X=X_{mb+}\doteq 2.38445$. 
The curve has two local maxima of the same value, $E_{mb}\doteq 0.93856$, corresponding to the inner 
[mb(i)] and the outer [mb(o)] marginally bound orbits. The dashed effective 
potential defines the inner marginally stable orbit [ms(i)] by coalescing the local minimum and 
the (inner) local maximum. It corresponds to the parameter $X=X_{ms(i)+}\doteq
2.20307$ with specific
energy $E_{ms(i)+}\doteq 0.90654$. In an analogous manner, the dash-dotted potential defines the  
outer marginally stable orbit [ms(o)] with specific energy $E_{ms(o)+}\doteq 0.94451$ corresponding to 
the parameter $X=X_{ms(o)+}\doteq 2.90538$.}
\label{f10}
\end{figure}
\begin{figure*}
\epsfxsize=.95 \hsize
\epsfbox{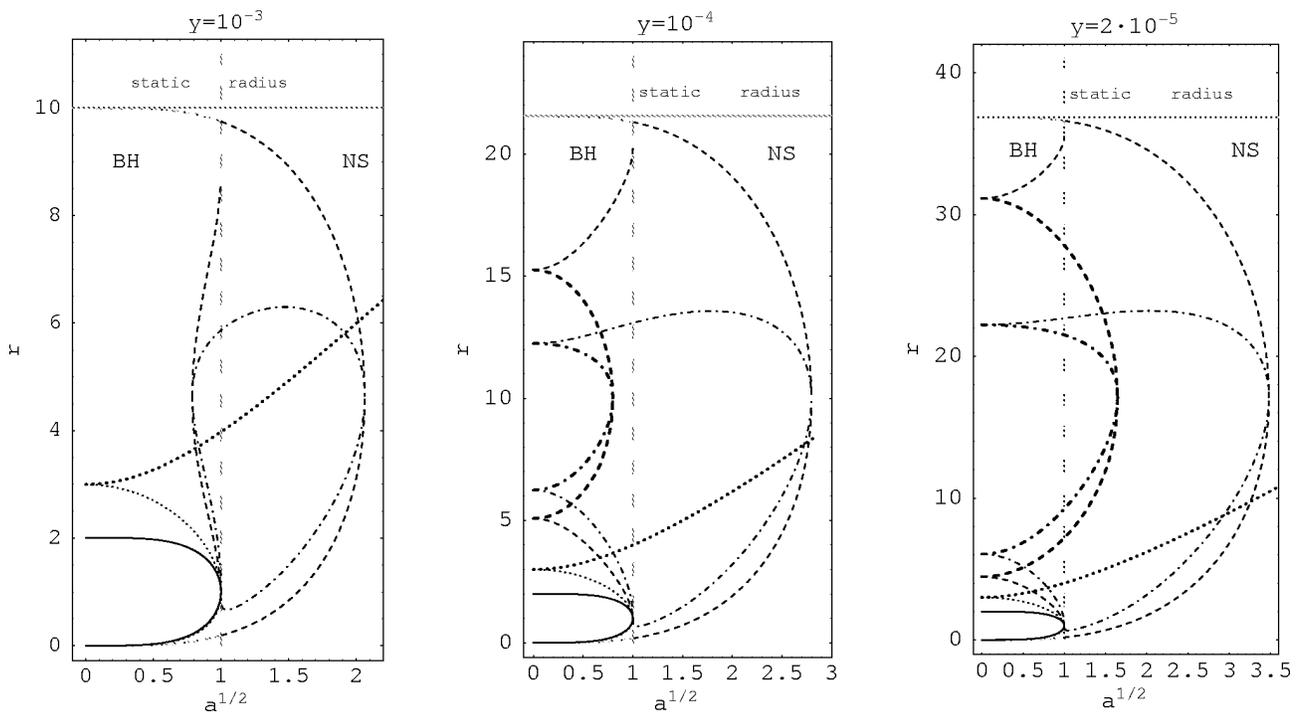}
\caption{Mutual positions of the astrophysically important circular orbits in Kerr--de~Sitter 
spacetimes. The figures are constructed for three representative values of $y$. The radii of the 
special equatorial circular orbits are plotted as functions of the rotational parameter $a$. 
The wide dashed line is given by the value of rotational parameter corresponding to the extreme black 
hole and it splits up Kerr--de~Sitter spacetimes into black-hole (BH) and naked-singularity 
(NS) regions. Thin curves are used for the plus-family orbits (in the most cases they correspond to the
corotating orbits from the point of view of the locally nonrotating observers, but there are 
exceptions described in the text). Bold curves are used for the minus-family orbits (in all spacetimes
under consideration:
counterrotating orbits). Solid curves determine the inner and outer black-hole horizons. Dotted 
curves determine the photon circular orbits; dashed curves determine the marginally bound (mb) circular 
orbits. There is a disconnection between BH and NS regions for the plus-family orbits. Lower gray 
dashed curves determine the marginally bound orbits hidden under the inner black-hole horizon; the 
upper one, approaching the static radius for small $a$, is its outer analogy. Dash-dotted curves 
determine the marginally stable (ms) orbits. For $y\geq 12/15^4$ there are no minus-family mb and ms 
orbits.}
\label{f15}
\end{figure*}
The behavior of the effective potential (\ref{e28}) enables us to introduce the
notion of the marginally bound orbits---i.e., unstable circular orbits
where a small radial perturbation causes infall of a particle from the orbit
to the center or its escape to the cosmological horizon. For
some special value of the axial parameter $X$, denoted as $X_{mb}$, the
effective potential has two local maxima related by the condition
\be
   E_{(+)}(r_{mb(i)};X_{mb},a,y)=E_{(+)}(r_{mb(o)};X_{mb},a,y),
\ee
and corresponding to both the inner and outer marginally bound orbits; see
Fig.~\ref{f10} (and Fig.~\ref{f14}, below). 
For completeness, the figures include the effective
potentials defining both the inner and outer marginally stable orbits
(corresponding to special values of the parameter $X$: $X_{ms(i)},\ X_{ms(o)}$). 
The search for the marginally bound orbits in a concrete \Kds spacetime must be
realized in a numerical way and can be successful only in the 
spacetimes admitting stable circular orbits. Clearly, in the spacetimes with
$y \geq 12/15^4$, the minus-family marginally bound orbits do not
exist. Figure~\ref{f3} offers insight into the
possibility of the existence of both stable and bound circular orbits of both
families. The limiting (solid) curves are obtained from the conditions
(\ref{e58}), (\ref{e64}) that have to be solved simultaneously.

The location of the astrophysically important circular orbits (photon
orbits, marginally stable and marginally bound orbits) in dependence
on the rotational parameter $a$ is given in Fig. \ref{f15} for three
appropriately chosen values of the cosmological parameter $y$. The
values of $y$ reflect the dependence of the existence of stable
minus-family orbits on $y$. Stable plus-family orbits exist for
all chosen values of $y$ in the relevant range of the parameter
$a$. Spacetimes without stable circular orbits or without any circular orbits are
inferred from Figs. \ref{f7}, \ref{f8}.

\section{Discussion}
\label{Dis}

\begin{figure}
\epsfxsize=0.95\hsize
\epsfbox{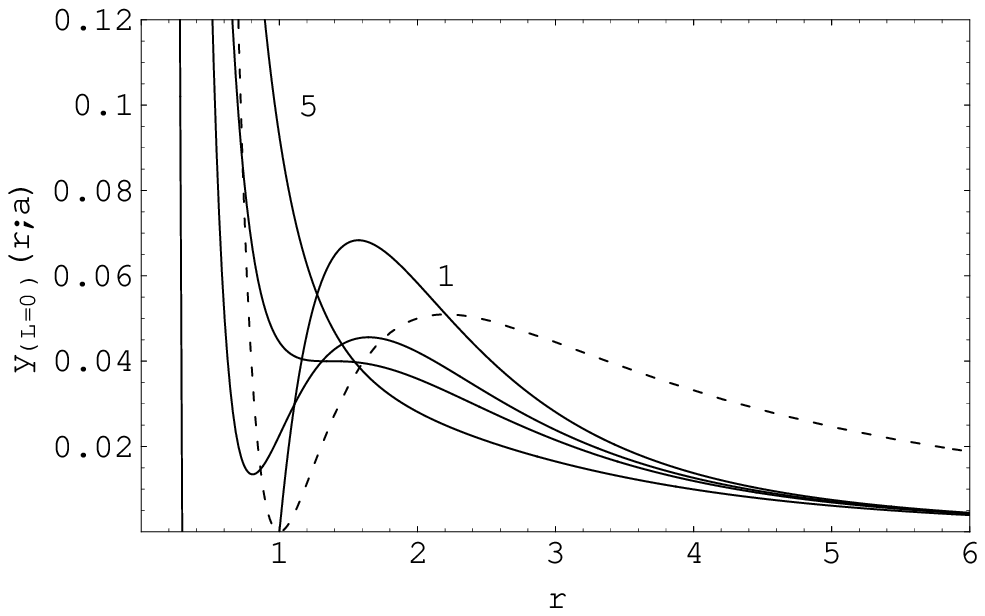}
\caption{Circular orbits with zero angular momentum. The function $y_{(L=0)}(r;a)$ is plotted for 
$a^2=\{1, 1.8, 2.4406, 5\}$. For $a^2=1$, when the black-hole spacetimes are allowed, the first two 
orbits are hidden under the black-hole event horizon. The dashed curve determines horizons of 
\Kds spacetimes with $a^2=1$. In the other cases only naked-singularity spacetimes are allowed and 
even three $L = 0$ orbits are possible for $a^2 < 2.4406$. The case $a^2\doteq 2.4406$ (the curve 
with an inflex point at $y\doteq 0.03998$) corresponds to the maximum value of the rotational 
parameter addmiting the stable counterrotating plus-family orbits.} 
\label{f11}
\end{figure}
In comparison with asymptotically flat Kerr spacetimes, where the effect of the rotational parameter
vanishes for asymptotically large values of the radius, in Kerr--de~Sitter spacetimes the properties 
of the circular orbits must be treated more carefully, because the rotational effect is relevant in 
whole the region where the circular orbits are allowed and it survives even at the cosmological horizon.

The minus-family orbits have specific angular momentum negative, $L_-<0$, in every Kerr--de~Sitter 
spacetime and such orbits are counterrotating from the point of view of locally nonrotating frames.

In black-hole spacetimes, the plus-family orbits are corotating in almost all radii where the circular 
orbits are allowed except some region in vicinity of the static radius, where they become 
counterrotating, as their specific angular momentum $L$ is slightly negative
there. In naked-singularity spacetimes, the plus-family orbits behave in
a more complex way; nevertheless, they are always counterrotating in vicinity
of the static radius.

The specific
angular momentum of particles located on the static radius, where the plus-family orbits and the 
minus-family orbits coalesce, is given by the relation
\begin{equation}
L(r_{s};y,a)=L_{s} \equiv -a\frac{3y^{1/3}+a^{2}y}{\left (1-3y^{1/3}-a^{2}y \right )^{1/2}},
\end{equation}
and their specific energy is
\be
E(r_s;y,a)=E_s\equiv (1-3y^{1/3}-a^2 y)^{1/2}.
\ee

\subsection{Circular orbits with zero angular momentum}

Separation of the corotating and counterrotating orbits as defined by their azimuthal 
angular momentum relative to the locally nonrotating frames is determined by the orbits with $L=0$. 
The orbits with zero angular momentum are defined by the relation
\begin{widetext}
\begin{equation}
y=y_{(L=0)}(r;a) \equiv \frac{-r[r(r^2+a^2)+4a^2]+r^{1/2}(r^2+a^2)^{1/2}\left [(r^2+a^2)(r^3+4a^2)+8a^2r^2\right ]^{1/2}}
{2a^2r^2(r^2+a^2)} \label{e79}.
\end{equation}
\end{widetext}
At these orbits, the locally nonrotating observers follow circular geodesics at the equatorial plane.

The physically relevant zero points of the function $y_{(L=0)}$ are given by the function 
\be
a_{z(L=0)}(r)\equiv r^{1/2}[1+(1-r)^{1/2}]
\ee
determining circular orbits with $L=0$ in the asymptotically flat Kerr backgrounds. A detailed study 
reveals that such orbits exist only in Kerr naked-singularity spacetimes with 
$1 < a/M \leq \frac{3}{4}\sqrt{3}$. (In Kerr black-hole spacetimes orbits of this kind are hidden 
under the event horizon.) The typical behavior of the function $y_{(L=0)}(r;a)$ is presented in 
Fig.~\ref{f11} for some appropriately chosen values 
of the rotational parameter $a^2$. 
The function has a local extremum for $a^2 < a^2_{cL} \equiv 2.4406$ 
(where the critical value $a^2_{cL}$ is obtained by a numerical procedure)---we can conclude that up to 
three zero-angular-momentum orbits can exist in the corresponding \Kds spacetimes. In the naked-singularity 
spacetimes all three orbits with zero angular momentum are relevant and the middle orbit is stable. 
In black-hole spacetimes, however, two of these orbits are hidden under the black-hole horizons and
only the unstable one, located nearby the static radius, is physically important. 
The plus-family orbits between the zero-angular-momentum orbit and the static radius are counterrotating. 
Comparison of the functions $y_{(L=0)}(r;a)$ and $y_{ms}(r;a)$ (determining the position of the 
outermost marginally stable orbit for a given cosmological parameter $y$) reveals that the discussed orbits 
are unstable.

In naked-singularity spacetimes, the behavior of the plus-family orbits
is more intriguing. Except for the unstable counterrotating orbits located nearby the static radius 
(discussed above), some stable counterrotating plus-family circular orbits exist in the vicinity of 
the ring singularity of naked-singularity spacetimes with rotational parameter low enough. 
Spacetimes admitting such orbits belong to the dash-dotted naked-singularity region of the parametric 
space $(y,a^2)$ presented in Fig.~\ref{f3}. The limiting (dash-dotted) curve was obtained by
solving simultaneously the conditions for the marginally stable orbits given
by Eq.~(\ref{e58}) and the condition for the orbits with zero angular momentum given by Eq.~(\ref{e79}).

In a given Kerr--de~Sitter naked-singularity spacetime relations (\ref{e58}), (\ref{e79}) determine 
the innermost and outermost stable counterrotating plus-family orbits, respectively.

\subsection{Circular orbits with negative energy}

\begin{figure*}
\epsfxsize=0.8\hsize
\epsfbox{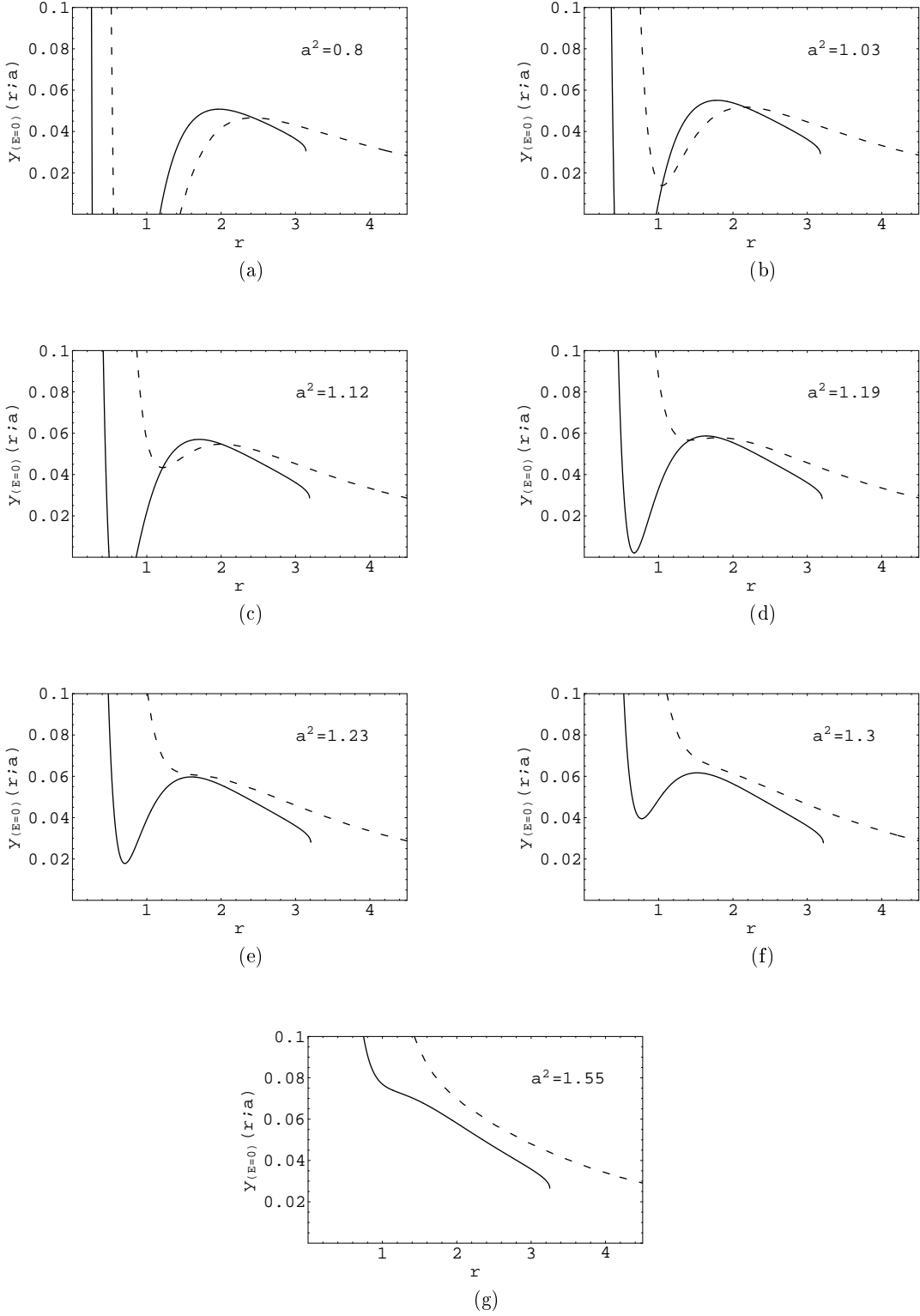}
\caption{Location of the plus-family orbits with $E=0$. The dashed curves
  determine the horizons of \Kds spacetimes. The solid curves determine
  the orbits with $E=0$. The rising (descending) part of $y_{(E=0)}(r;a)$
  corresponds to the stable (unstable) orbits. The intervals of the
  rotational parameter giving different behavior of the curves
  $y_{(E=0)}(r;a)$, $y_h (r;a)$ in cases (a)--(g) are (a) $0<a^2<1$, (b)
  $1<a^2<1.06992$, (c) $1.06992<a^2<a^{2}_{c(E=0)}\doteq 1.18518$, (d)
  $a^{2}_{c(E=0)}<a^2<a^{2}_{\rm {crit}}\doteq 1.21202$, (e) $a^{2}_{\rm
  {crit}}<a^2<1.25976$, (f) $1.25976<a^2<a^2_{s(E=0)}\doteq 1.47000$, (g) $a^2>a^2_{s(E=0)}$.}
\label{f12}
\end{figure*}

In the rotating naked-singularity spacetimes the potential well can be deep enough nearby the ring 
singularity to allow the existence of stable (plus-family) counterrotating circular orbits with
negative specific energy, indicating an extremely high efficiency of conversion of the rest mass into 
heat energy during accretion in a corotating (or, more precisely, a plus-family) thin disk.
The plus-family circular orbits with zero energy are given by the relation
\begin{widetext}
\bea
     y=y_{(E=0)}(r;a)\equiv
     \frac{r[2(r^2+a^2)(r-2)-a^2 r]+ ar^{1/2}
     \{4(r^2+a^2)^2-r^2[4(r^2+a^2)(r-2)-a^2r]\}^{1/2}}{2r^2(r^2+a^2)^2}.
                                              \label{e82}
\eea
\end{widetext}
The reality conditions of the function $y_{(E=0)}(r;a)$ are given by the relations
\bea
r\geq 0, \label{e83}\\
4(r^2+a^2)^2-r^2[4(r^2+a^2)(r-2)-a^2r]\geq 0. \label{e84}
\eea
The condition (\ref{e84}) can be transferred into the relation 
\be\label{e85}
a^2\geq a^2_{min(E=0)}(r)\equiv \frac{r^2}{8}[3r-16+(9r^2-32r+64)^{1/2}],
\ee
which is relevant for $r>3$. For $0<r<3$ the function  $a^2_{min(E=0)}(r)$ is negative.
The zero points of the function $y_{(E=0)}(r;a)$ are given by the function
\be\label{e86}
                  a_{z(E=0)}(r)\equiv r^{1/2}(2-r),
\ee
which determines the circular orbits with zero specific energy in Kerr spacetimes. For $y=0$, such 
orbits exist only in Kerr naked-singularity  spacetimes with 
$1 < a/M \leq \frac{4}{3}\sqrt\frac{2}{3}$, which are a subset of spacetimes with zero-angular-momentum 
orbits; in fact, orbits with $E=0$ have $L<0$ (for details see \cite{Stu:1980:BULAI:}). 
The behavior of the function $y_{(E=0)}(r;a)$ is presented for some typical values of the rotational 
parameter $a$ in Fig.~\ref{f12}. 
\begin{figure*}
\epsfxsize=1 \hsize
\epsfbox{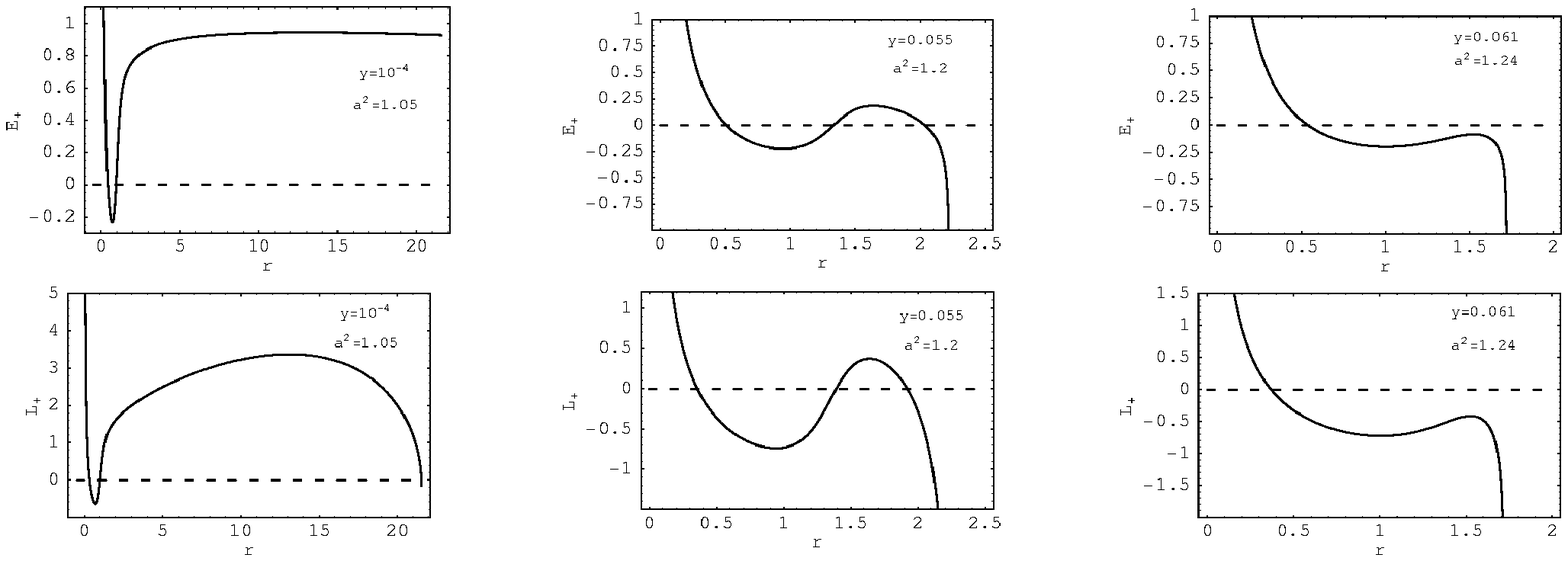}
\caption{Naked-singularity spacetimes admitting stable circular orbits with negative specific energy. 
We can see that stable circular orbits in vicinity of the innermost stable orbit (the left
and middle columns) or even all the 
stable orbits (the right column) can have negative energy. The behavior of the specific angular 
momentum reveals counterrotation of such orbits.}
\label{f13}
\end{figure*}
In \Kds spacetimes with $0<a^2<1.47$, the function $y_{(E=0)}$ has 
two local extrema leading up to three circular orbits with zero energy. The ending points of the curves 
are  given by the condition (\ref{e85}) and are represented by the function 
\bea
     \lefteqn{y_{min(E=0)}(r)\equiv y_{(E=0)}(r;a=a_{min(E=0)})=} \nonumber \\
          & & \frac{3r^2-10r+16-(r-2)(9r^2-32r+64)^{1/2}}{2r^4}. \nonumber \\
                                                             \label{e87}
\eea
Details of the properties of the plus-family orbits with $E=0$ can be inferred
from Fig.~\ref{f12}. Here, we give a short overview of them.

In the black-hole spacetimes, there is always one orbit with $E=0$ located under the
inner black-hole horizon, and there can exist, for properly chosen parameters $a$
and $y$, one orbit with $E=0$ located above the outer black-hole horizon. Both
the orbits must be unstable relative to radial perturbations.

In the naked-singularity spacetimes,  
if $a^2 < a^{2}_{c(E=0)}\doteq 1.18518$, there can exist one orbit with $E=0$ (unstable),
two such orbits (the inner one unstable, the outer one stable), or three such
orbits (the inner and outer being unstable, the intermediate being stable). 
If $a^2 > a^{2}_{c(E=0)}$ and $y$ is properly chosen, there can be an
additional possibility of the nonexistence of the circular orbit with
$E=0$. If $a^2 > a^{2}_{s(E=0)}\doteq 1.47000$, there can exist no stable
zero-energy orbits for any $y$ (cf. Fig.~\ref{f3}).

\begin{figure}
\epsfxsize=0.99\hsize
\epsfbox{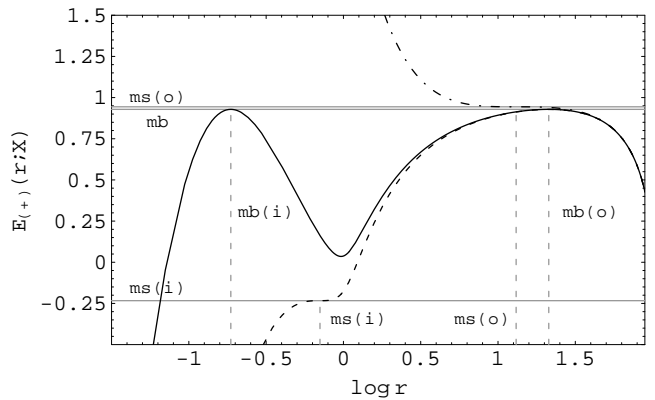}
\caption{Effective potential of the equatorial radial motion of test particles in an appropriately 
chosen \Kds naked-singularity spacetime ($y=10^{-4},\ a^2=1.05$) allowing the stable circular orbits 
with negative energy. Marginally bound (mb) orbits are given by the solid curve corresponding to 
$X=X_{mb+}\doteq -0.18444$. The curve has two local maxima of the same value, $E_{mb}\doteq 0.92857$, 
corresponding to the inner [mb(i)] and the outer [mb(o)] marginally bound orbits. The dashed effective 
potential defines the inner marginally stable orbit [ms(i)] resulting from coalescence of the local 
minimum and (inner) local maximum, with $X=X_{ms(i)+}\doteq -0.40713$ and the specific energy 
$E_{ms(i)+}\doteq -0.23321$. In an analogous manner, the dash-dotted potential defines the outer 
marginally stable orbit [ms(o)] with the specific energy $E_{ms(o)+}\doteq 0.94405$ and 
$X=X_{ms(o)+}\doteq 2.40379$.}
\label{f14}
\end{figure}

Examples of naked-singularity spacetimes admitting stable counterrotating plus-family circular 
orbits with negative energy are presented in Fig.~\ref{f13}.
The efficiency of conversion of the rest mass 
into heat energy during accretion, given by the relation 
\be
            \eta \equiv E_{ms(o)} - E_{ms(i)},      \label{e88}
\ee
is limited by the specific energy of the outermost stable circular plus-family orbit,
$E_{ms(o)}<1$, which can be directly inferred from Fig.~\ref{f14}.
Correspondingly, extraction of the rotational energy from a naked singularity with rotational parameter 
low enough is possible with subsequent conversion of the naked singularity into a black hole 
(see, e.g., \cite{Stu:1981:BULAI:}). 
Spacetimes allowing such processes are contained in the shaded naked-singularity region of the parametric 
space $(y,a^2)$ in the Fig.~\ref{f3}. 
The limiting (dotted) curve was obtained by solving simultaneously conditions for the marginally 
stable orbits (\ref{e58}) and the circular orbits with $E=0$
(\ref{e82}).

\section{Concluding remarks}
\label{Conclusion}

\begin{figure*}
\epsfxsize=1\hsize
\epsfbox{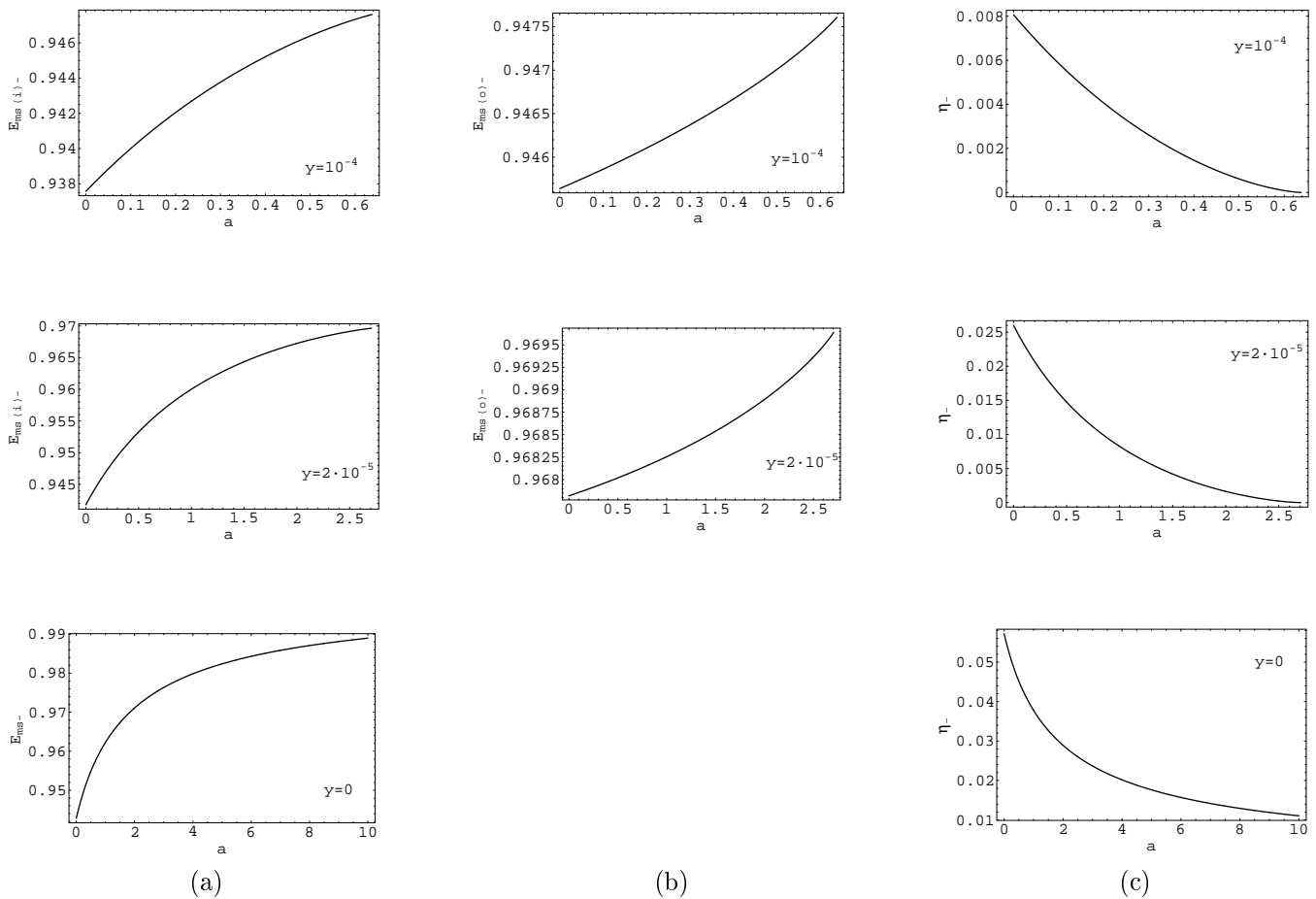}
\caption{Specific energy of the marginally stable minus-family orbits
  [{\bf (a)} Inner, 
{\bf (b)} Outer] and {\bf (c)} the accretion 
efficiency $\eta_{-} \equiv E_{ms(o)-}-E_{ms(i)-}$ given as
a function of the rotational parameter for three representative values of the
cosmological parameter. For Kerr spacetimes, $y=0$, we assume $E_{ms(o)-}=1$.}
\label{f16}
\end{figure*}
Many properties of Kerr--de Sitter spacetimes and circular orbits of
both families can be clearly viewed
from figures which are presented in the paper. Table \ref{t1} contains a
certain classification of the figures which could be helpful for quick
orientation in the topic.
\begin{table}[h]
\caption{Classification of the figures.}
\vspace{10pt}
\begin{tabular}{lll}
\hline\hline
Spacetime properties & Figs. 1--3 & \\
\hline
Properties & $(+)$ family & $(-)$ family \\ 
of circular orbits & & \\
\hline
\hspace*{5pt} General & Figs. 7--11, 15 & Figs. 7--9, 11 \\
\hspace*{5pt} Specific energy & Figs. 4, 6, 13, 14, 17--19 &
Figs. 4, 6, 16 \\
\hspace*{5pt} Spec. ang. mom. & Figs. 5, 6, 12, 19 &
Figs. 5, 6, 12 \\
\hspace*{5pt} Accretion efficiency & Figs. 17, 18 & Fig. 16 \\
\hline\hline
\end{tabular}
\label{t1}
\end{table}
\begin{figure*}
\epsfxsize=1 \hsize
\epsfbox{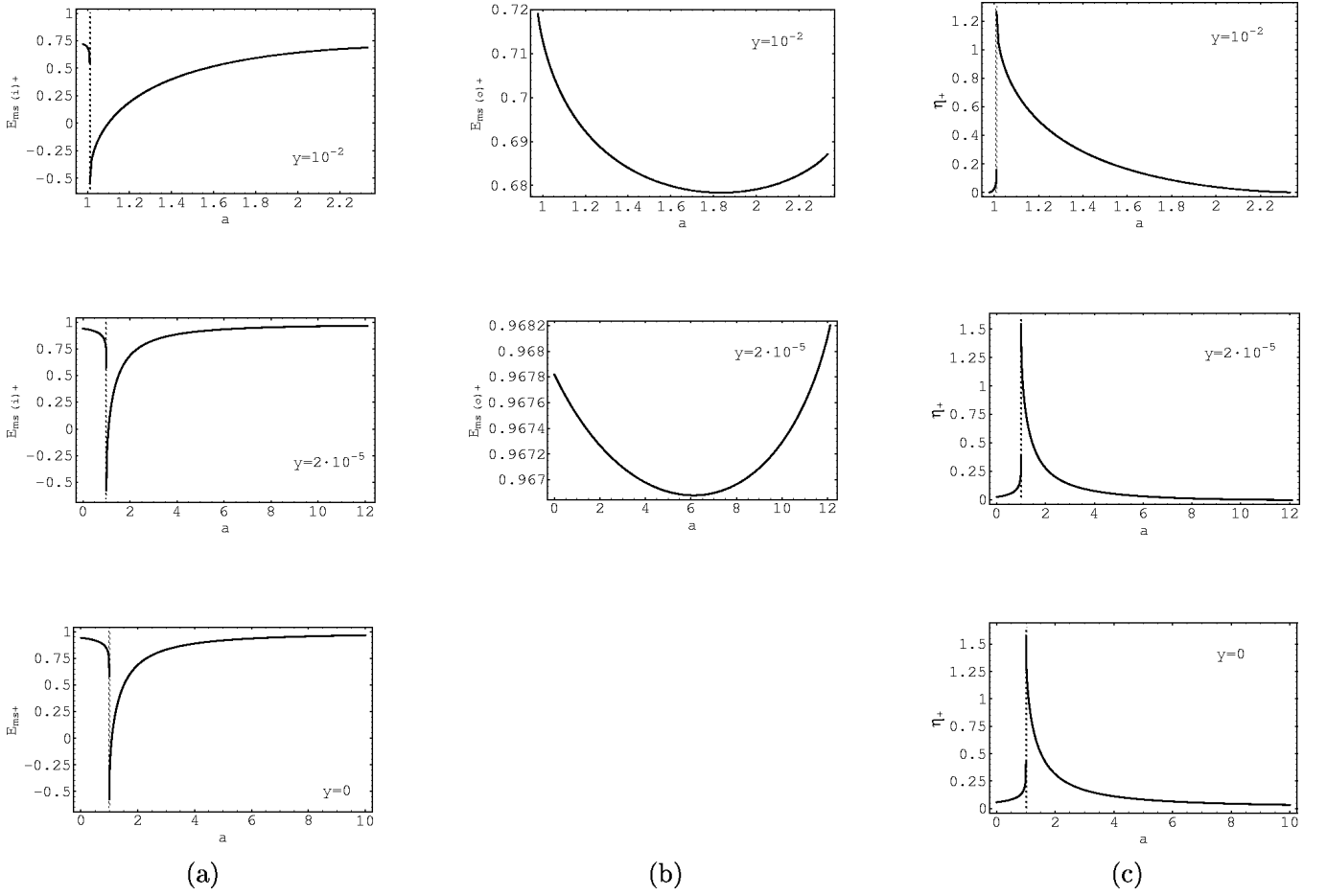}
\caption{Specific energy of the marginally stable plus-family orbits [{\bf (a)} Inner, 
{\bf (b)} Outer] and {\bf (c)} the accretion efficiency $\eta_{+} \equiv E_{ms(o)+}-E_{ms(i)+}$ 
(right column) as a function of the rotational parameter for three representative values of the
cosmological parameter. The gray line corresponds to the extreme black hole. We can see a strong 
discontinuity of the specific energy of the inner marginally stable orbits and the accretion
efficiency when black holes and naked singularities approach the extreme
black-hole state. For Kerr spacetimes, $y=0$, we assume $E_{ms(o)+}=1$.}  
\label{f17}
\end{figure*}

Both black-hole and naked-singularity Kerr--de~Sitter spacetimes can be separated into three classes 
according to the existence of stable (and, equivalently, marginally bound) circular orbits (see 
Fig.~\ref{f3}). Stable orbits of both the plus family and minus family exist in the spacetimes of 
class I (black holes) and class V (naked singularities). Solely stable orbits of the plus family exist 
in the spacetimes of classes II (black holes) and VI (naked singularities). No stable orbits exist 
in the spacetimes of classes III and IV. In dependence on the cosmological parameter, there are three 
qualitatively different types of the behavior of the loci of the marginally stable, marginally bound, 
and photon circular orbits as functions of the rotational parameter. These functions are illustrated for 
three representative values of $y$ in Fig.~\ref{f15}, enabling us to make in a straightforward way separation 
of Kerr--de~Sitter spacetimes into classes I--VI. 
In the special case of Kerr spacetimes 
($y=0$), these functions can be found in \cite{Bar:1973:BlaHol:,%
                                               Stu:1980:BULAI:}.  

The marginally stable circular orbits are crucial in the context of Keplerian (geometrically thin) 
accretion disks as these orbits determine the efficiency of conversion of rest mass into heat energy 
of any element of matter transversing the disks from their outer edge located on the outer marginally stable 
orbit to their inner edge located on the inner marginally stable orbit. 

Clearly, accretion disks constituted from minus-family orbits are everywhere counterrotating 
relative to the locally nonrotating frames. For the minus-family disks, the specific energy of both the 
outer and inner marginally stable circular orbits and the efficiency parameter 
$\eta_{-} = E_{ms(o)-} - E_{ms(i)-}$ are given for three typical values of $y$ as functions of $a$ in 
Fig.~\ref{f16}.
In the limit of $a \rightarrow 0$ with $y$ being fixed, we obtain the known values of 
the specific energy $E_{ms(o)}$, $E_{ms(i)}$ and the efficiency parameter of the accretion process 
$\eta$ for the Schwarzschild--de~Sitter black holes \cite{Stu-Hle:1999:PHYSR4:}. 
Both the specific energy parameters $E_{ms(o)-}(a)$, $E_{ms(i)-}(a)$ and the efficiency $\eta_{-}(a)$ 
vary smoothly at values of the rotational parameter corresponding to the extreme black holes. 

The Keplerian accretion disks constituted from the plus-family orbits behave in much more complex way in 
comparison with those of the minus-family orbits. First, usually these disks could be considered as 
corotating relative to the locally nonrotating frames; recall that in asymptotically flat Kerr 
black-hole spacetimes the plus-family disks are corotating at all radii down to the marginally 
stable orbit, while in the field of naked singularities with $a/M < \frac{3}{4}\sqrt{3}$ the stable 
circular orbits corotating at large distances are transformed into counterrotating orbits in 
vicinity of the marginally stable orbit \cite{Stu:1980:BULAI:}. A similar behavior occurs in 
Kerr--de~Sitter spacetimes; however, in the spacetimes with $y \rightarrow y_{c(KdS)}$, the stable 
plus-family orbits can be counterrotating even at all allowed radii (see, e.g., Fig.~\ref{f13}). 
(Moreover, there are always counterrotating plus-family orbits in the vicinity of the static radius, 
where the plus-family orbits and the minus-family orbits coalesce; these orbits are, however, unstable 
relative to radial perturbations and cannot be related to accretion disks.) 

\begin{figure*}
\epsfxsize=1 \hsize
\epsfbox{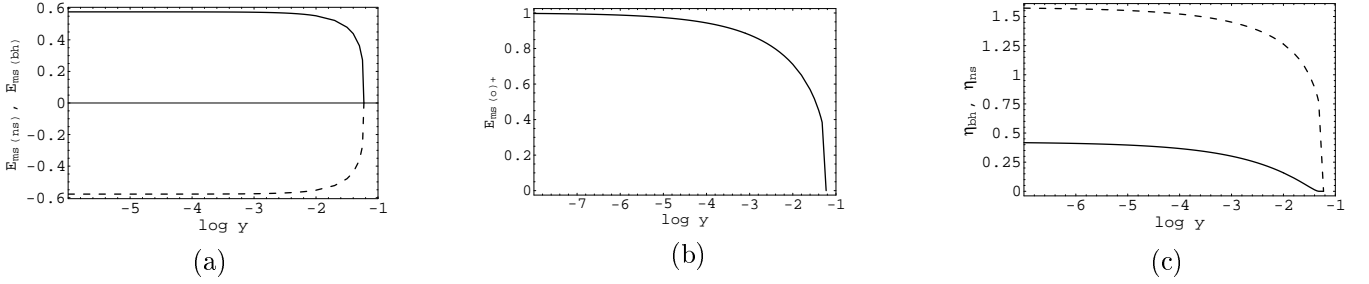}
\caption{Specific energy of the marginally stable orbits and accretion efficiency near the 
extreme black-hole states. {\bf (a)} Specific energy of the inner marginally stable plus-family 
orbit in extreme black-hole and related limiting naked-singularity spacetimes approaching the 
extreme hole states as a function of the cosmological parameter $y$. The solid curve corresponds to the 
extreme black holes; the dashed curve corresponds to the limiting naked singularities. The curves are 
symmetric around the zero-energy axis and tend to zero for $y=y_{c(KdS)}$. In extreme Kerr 
spacetimes ($y=0$), the specific energies in the black-hole and naked-singularity cases are 
$1/\sqrt{3}$ and $-1/\sqrt{3}$, respectively. {\bf (b)} Specific energy of the outer marginally stable 
plus-family orbit in extreme Kerr--de~Sitter black-hole spacetimes is the same as for 
naked-singularity spacetimes approaching the extreme hole state; i.e., there is no discontinuity in 
this case. The specific energy tends to zero for $y\to y_{c(KdS)}$. {\bf (c)} Accretion 
efficiency for the extreme black holes $\eta_{bh}$ (solid curve) and for the limiting naked 
singularities $\eta_{ns}$ (dashed curve). For $y=0$ (pure Kerr spacetimes) we obtain 
the maximum value $0.42$ for black holes and $1.58$ for naked singularities. For $y\to y_{c(KdS)}$ 
the efficiency tends to zero for both black holes and naked singularities.}
\label{f18}
\end{figure*}
\begin{figure*}
\epsfxsize=1\hsize
\epsfbox{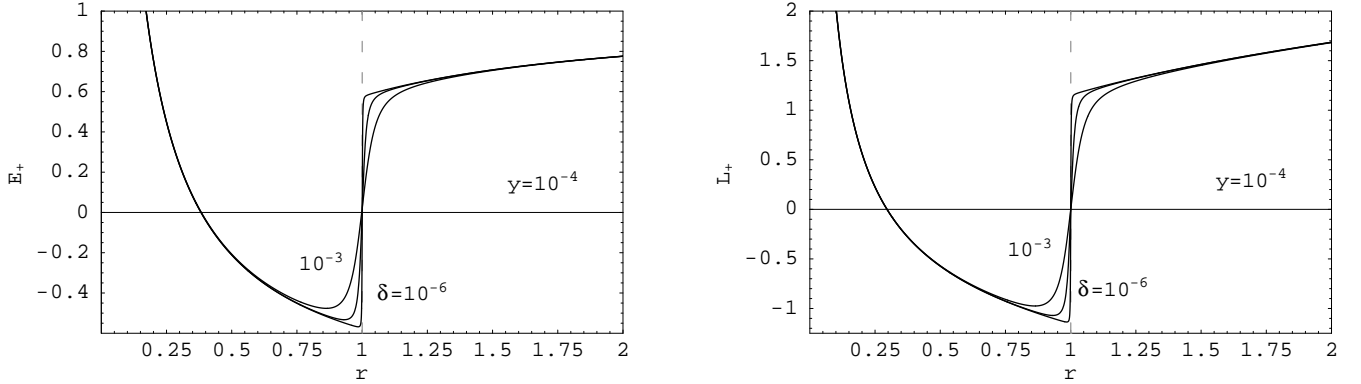}
\caption{Distribution of the specific energy and the specific angular momentum of the equatorial 
circular orbits in naked-singularity spacetimes approaching the extreme black-hole state. The 
orbits with the specific energy in the interval $E_{ms(ns)} < E < E_{ms(bh)}$ are located in an 
extremely small interval of the radial coordinate having, however, an extremely long proper length 
\protect\cite{Bar:1973:BlaHol:}. After conversion of a hypothetical naked singularity into an extreme 
black hole all these circular orbits become unstable relative to radial perturbations and will be
immediately swallowed by the black hole. The figures are drawn for $a = a_{0}(1+\delta)$ and
$y = y_{0}(1-\delta)$, where $y_0 = 10^{-4}$ and $a_0 = 1.0001$ are chosen to correspond to an
extreme black hole and, subsequently, $\delta = 10^{-3}, 10^{-4}, 10^{-6}$.} 
\label{f19}
\end{figure*}
Second, the specific energy $E_{ms(i)+}(y,a)$ of the inner marginally stable plus-family orbit can be 
negative. Recall that $E_{ms(i)+} < 0$ in asymptotically flat Kerr naked-singularity spacetimes 
with the rotational parameter $a/M < \frac{4}{3}\sqrt{\frac{2}{3}}$, indicating the efficiency of the 
accretion process $\eta_{+} = E_{ms(o)+} - E_{ms(i)+} > 1$, because in asymptotically flat 
Kerr spacetimes the outer edge of the accretion disks can be at arbitrarily large radii, implying 
thus $E_{ms(o)+} = 1$. In 
Kerr--de~Sitter spacetimes allowing $E_{ms(i)+} < 0$, the efficiency of the accretion process can be 
both $\eta > 1$ and $\eta < 1$, as it depends strongly on $E_{ms(o)+}$, which for $y \sim y_{c(KdS)}$ 
can be even negative (see Fig.~\ref{f13}). For three typical values of $y$, the functions $E_{ms(o)+}(a)$, 
$E_{ms(i)+}(a)$, $\eta_{+}(a)$ are illustrated in Fig.~\ref{f17}.
The specific energy function 
$E_{ms(i)+}(a)$ falls for $a$ growing in the black-hole region and for $a$ descending in the 
naked-singularity region. The specific energy function $E_{ms(o)+}(a)$ has a local minimum at some 
value of the rotational parameter $a$ strongly dependent on the cosmological parameter $y$. For $y$ 
being fixed, the accretion efficiency $\eta_{+}(a)$ grows for $a$ growing in the black-hole sector up 
to the critical value corresponding to the extreme black-hole spacetime, and it also grows for $a$ 
descending in the naked-singularity sector down to the critical value. 

Third, there is a strong discontinuity of the specific energy function $E_{ms(i)+}(a)$ for spacetimes 
approaching the extreme black hole state from the black-hole and the naked-singularity sectors. For 
extreme Kerr black holes ($y=0, a/M=1$), there is the limiting value of the specific energy 
$E_{ms(bh)} = 1/\sqrt{3}$, while for naked singularities approaching the extreme hole states 
($a/M \rightarrow 1$ from above), there is $E_{ms(ns)} = -1/\sqrt{3}$. For extreme 
Kerr--de~Sitter spacetimes, the dependence of the specific energy of the inner marginally stable orbit 
on the cosmological parameter is shown in Fig.~\ref{f18}a. Clearly, there is 
$E_{ms(ns)}(y) = - E_{ms(bh)}(y)$, where for a given 
cosmological parameter $y$ the rotational parameter $a$ of the corresponding extreme black hole is 
determined by the upper branch of the limiting line separating black-hole and naked-singularity states 
in Fig.~\ref{f3}. For $y \rightarrow y_{c(KdS)}$, there is $E_{ms(bh)}(y) \rightarrow 0$. For the 
specific energy function $E_{ms(o)+}(y,a)$ of the outer marginally stable orbits there is no 
discontinuity at the states corresponding to extreme black-hole spacetimes (see Fig.~\ref{f18}b). 
The accretion efficiency $\eta_{+}(y)$ in the field of extreme black holes [$\eta_{bh}(y)$] and in the 
field of the naked singularities infinitesimally close to the extreme hole states [$\eta_{ns}(y)$] is 
shown in Fig.~\ref{f18}c.
For $y = 0$ their difference takes the maximum 
($\eta_{ns} = 1 + 1/\sqrt{3}$, $\eta_{bh} = 1 - 1/\sqrt{3}$), while at 
$y = y_{c(KdS)}$ the difference vanishes ($\eta_{ns} = 0$, $\eta_{bh} = 0$). 

As a result of accretion in a plus-family or a minus-family Keplerian disk, a hypothetical naked singularity 
can be converted into an extreme black hole. In the case of Kerr naked singularities their evolution 
into an extreme hole state was discussed in \cite{Cal-Nob:1979:NUOC2:,%
                                                  Stu:1981:BULAI:,%
                                                  Stu-Pls-Hle:2002:PERSEUS:EvoKerrNakSin}. 
Such a 
conversion can be a rather dramatic process in the case of the plus-family accretion disks, because 
of the discontinuity of the plus-family orbits at the extreme black-hole state. We can understand 
this process if we show how the stable circular orbits are distributed in naked-singularity 
spacetimes approaching the extreme black-hole state (Fig.~\ref{f19}).
We can see that all the orbits 
with the specific energy ranging from $E_{ms(ns)}(y)$ up to $E_{ms(bh)}(y)$ are distributed at an 
infinitesimally small range of the radial coordinate in the vicinity of the radius corresponding to the 
event horizon of the extreme black hole. Of course, it is well known that at these radii the physically 
relevant proper radial length, along which the accretion disk is distributed, becomes very (almost 
infinitely) long (see \cite{Bar:1973:BlaHol:}). If the conversion of a hypothetical naked 
singularity into an extreme black hole is realized, the part of the accretion disk located under 
the marginally stable circular orbit of the created black hole becomes unstable relative to radial 
perturbations and will be immediately swallowed by the black hole. It can be expected that the 
collapse of the unstable internal part of the disk with the specific energy ranging from 
$E_{ms(ns)}(y)$ up to $E_{ms(bh)}(y)$ could be observationally important, leading to an abrupt 
fall down of observable luminosity of the accretion disk.

\begin{table}
\caption{Mass parameter, static radius, and radius of the outer 
marginally stable circular orbit determining the outer edge of
corotating Keplerian disks in extreme Kerr--de Sitter black-hole
spacetimes are given for the relict repulsive cosmological constant
indicated by recent cosmological observations: $\Lambda_0\approx 1.1
\times 10^{-56}\ \rm{cm}^{-2}$. Note that accretion efficiency $\eta_+$ is,
in principle, smaller than the value for the pure Kerr case ($y=0$)
but, in practice, for small values of cosmological parameter $y$
contained in the table, $\eta_+$ is undistinguishable from the Kerr limit
$\eta\approx 0.42$.}
\vspace{10pt}
\begin{tabular}{cccc}
\hline\hline
$y$ & $M$ & $r_s$ & $r_{ms(o)+}$ \\
    & [$M_{\odot}$] & [kpc] & [kpc] \\
\hline
$10^{-46}$ & 1.1 & 0.1 & 0.07 \\
$10^{-44}$ & 11.1 & 0.2 & 0.15 \\
$10^{-42}$ & 111.4 & 0.5 & 0.3 \\
$10^{-40}$ & $1.1 \times 10^3$ & 1.1 & 0.7 \\
$10^{-34}$ & $1.1 \times 10^6$ & 11.4 & 7.2 \\
$10^{-32}$ & $1.1 \times 10^7$ & 24.5 & 15.5 \\
$10^{-30}$ & $1.1 \times 10^8$ & 52.8 & 33.3 \\
$10^{-28}$ & $1.1 \times 10^9$ & 113.8 & 71.7 \\
$10^{-26}$ & $1.1 \times 10^{10}$ & 245.2 & 154.5 \\
$10^{-24}$ & $1.1 \times 10^{11}$ & 528.3 & 332.9 \\
$10^{-22}$ & $1.1 \times 10^{12}$ & 1138.4 & 717.1 \\
\hline\hline
\end{tabular}
\label{t2}
\end{table}
Finally, we shall give to our results proper astrophysical relevance by
presenting numerical estimates for the observationally established value
of the current value of the cosmological constant. A wide range of
recent cosmological observations give a strong ``concordance''
indication \cite{Kra:1998:ASTRJ2:}
that the observed value of the vacuum energy density is
\be
      \varrho_{vac(0)}\approx 0.66 \varrho_{crit(0)},
\ee
with present values of the critical energy density
$\varrho_{crit(0)}$  and the Hubble parameter $H_{0}$ given by
\be
      \varrho_{crit(0)}=\frac{3H_{0}^2}{8\pi},\quad H_{0}=100h\
      \rm{km}\ \rm{s}^{-1}\ \rm{Mpc}^{-1}.
\ee
Taking the value of the dimensionless parameter $h\approx 0.7$, we obtain
the ``relict'' repulsive cosmological constant to be
\be
      \Lambda_{0}=8\pi\varrho_{vac(0)}\approx 1.1 \times 10^{-56}\ \rm{cm}^{-2}.
\ee
Having this value of $\Lambda_{0}$, we can determine the mass parameter
of the spacetime corresponding to any value of $y$, parameters of the
equatorial circular geodesics, and basic characteristics of the thin
accretion disks. For extreme black holes (we have chosen some typical values of
the black-hole mass), the dimensions of the static radius and the outer
marginally stable circular orbit
of the plus-family accretion disk are given in Table
\ref{t2}. For more detailed information in the case of thick disks
around \Sds black holes see \cite{Stu-Sla-Hle:2000:ASTRA:}, where the
estimates for primordial black holes in the early Universe with a
repulsive cosmological constant related to a hypothetical vacuum
energy density connected with the electroweak symmetry breaking or the
quark confinement are presented.

It is well known (see, e.g., \cite{Car-Ost:1996:ModAst:}) that 
dimensions of accretion disks around stellar-mass black holes ($M
\sim 10 M_\odot$) in binary systems are typically $10^{-3}$ pc, dimensions of
large galaxies with central black-hole mass $M \sim 10^8 M_\odot$, of
both spiral and elliptical type, are in the interval 50--100 kpc, and
extremely large elliptical galaxies of cD type with central
black-hole mass $M \sim 3\times 10^9 M_\odot$ extend up to 1
Mpc. Therefore, we can conclude that the influence of the relict
cosmological constant is quite negligible in the accretion disks in
binary systems of stellar-mass black holes as the static radius
exceeds in many orders dimension of the binary systems. But it can be
relevant for accretion disks in galaxies with large active nuclei as
the static radius puts limit on the extension of the disks well inside
of the galaxies. Moreover, the agreement (up to one order) of the
dimension of the static radius related to the mass parameter of
central black holes at nuclei of large or extremely large galaxies
with extension of such galaxies suggests that the relict cosmological
constant could play an important role in the formation and evolution of such
galaxies. Of course, the first step in confirming such a suggestion is
modeling of the influence of the repulsive cosmological constant on
self-gravitating accretion disks. Some hints this way could be
given by recent results of Rezzolla {\it et
al.} \cite{Rez-Zan-Fon:2003:ASTRA:}, based on sophisticated numerical
hydrodynamic methods developed by Font \cite{Fon-Dai:2002:MONNR:,%
                                             Fon-Dai:2002:ASTRJ2L:}, 
who showed that mass outflow from
the outer edge of thick accretion disks, induced by the relict
cosmological constant, could efficiently stabilize the accretion disks
against the runaway dynamical instability. 

\section*{ACKNOWLEDGMENTS}
The present work was supported by GA\v{C}R grant No.~205/03/1147 and by the Bergen Computational 
Physics Laboratory project, a EU Research Infrastructure at the University of Bergen, Norway,
supported by the European Community Access to Research Infrastructure Action of the Improving
Human Potential Programme. The authors would like to express their gratitude to Professor~L.~P.~Csernai
for hospitality at the University of Bergen. Z.S. and P.S. would
like to acknowledge the excellent working conditions at the CERN's Theory
Division and SISSA's Astrophysic Sector, respectively.


\end{document}